\documentclass[aps,groupedaddressamsmath,amssymb,
 prd, superscriptaddress,twocolumn]{revtex4-2}

\usepackage{graphicx}
\usepackage{dcolumn}
\usepackage{bm}
\usepackage{hyperref}
\usepackage[mathlines]{lineno}
\usepackage[normalem]{ulem}
\usepackage{soul}
\usepackage{placeins}
\usepackage[export]{adjustbox}
\usepackage{ amssymb }
\usepackage{rotating}
\usepackage{siunitx}

\usepackage{lineno}
\usepackage{tabularx}
\usepackage[utf8]{inputenc}
\usepackage[T1]{fontenc}

\hypersetup{
    pdfnewwindow=true,    % links in new window
    colorlinks=true,      % false: boxed links; true: colored links
    linkcolor=blue,       % color of internal links
    citecolor=blue,       % color of links to bibliography
    filecolor=blue,       % color of file links
    urlcolor=blue         % color of external links
}

\usepackage[usenames, dvipsnames]{color}

\begin{document}

%Title of paper
\title{Search for GeV-scale Dark Matter Annihilation in the Sun with IceCube DeepCore}
\affiliation{III. Physikalisches Institut, RWTH Aachen University, D-52056 Aachen, Germany}
\affiliation{Department of Physics, University of Adelaide, Adelaide, 5005, Australia}
\affiliation{Dept. of Physics and Astronomy, University of Alaska Anchorage, 3211 Providence Dr., Anchorage, AK 99508, USA}
\affiliation{Dept. of Physics, University of Texas at Arlington, 502 Yates St., Science Hall Rm 108, Box 19059, Arlington, TX 76019, USA}
\affiliation{CTSPS, Clark-Atlanta University, Atlanta, GA 30314, USA}
\affiliation{School of Physics and Center for Relativistic Astrophysics, Georgia Institute of Technology, Atlanta, GA 30332, USA}
\affiliation{Dept. of Physics, Southern University, Baton Rouge, LA 70813, USA}
\affiliation{Dept. of Physics, University of California, Berkeley, CA 94720, USA}
\affiliation{Lawrence Berkeley National Laboratory, Berkeley, CA 94720, USA}
\affiliation{Institut f{\"u}r Physik, Humboldt-Universit{\"a}t zu Berlin, D-12489 Berlin, Germany}
\affiliation{Fakult{\"a}t f{\"u}r Physik {\&} Astronomie, Ruhr-Universit{\"a}t Bochum, D-44780 Bochum, Germany}
\affiliation{Universit{\'e} Libre de Bruxelles, Science Faculty CP230, B-1050 Brussels, Belgium}
\affiliation{Vrije Universiteit Brussel (VUB), Dienst ELEM, B-1050 Brussels, Belgium}
\affiliation{Department of Physics and Laboratory for Particle Physics and Cosmology, Harvard University, Cambridge, MA 02138, USA}
\affiliation{Dept. of Physics, Massachusetts Institute of Technology, Cambridge, MA 02139, USA}
\affiliation{Dept. of Physics and Institute for Global Prominent Research, Chiba University, Chiba 263-8522, Japan}
\affiliation{Department of Physics, Loyola University Chicago, Chicago, IL 60660, USA}
\affiliation{Dept. of Physics and Astronomy, University of Canterbury, Private Bag 4800, Christchurch, New Zealand}
\affiliation{Dept. of Physics, University of Maryland, College Park, MD 20742, USA}
\affiliation{Dept. of Astronomy, Ohio State University, Columbus, OH 43210, USA}
\affiliation{Dept. of Physics and Center for Cosmology and Astro-Particle Physics, Ohio State University, Columbus, OH 43210, USA}
\affiliation{Niels Bohr Institute, University of Copenhagen, DK-2100 Copenhagen, Denmark}
\affiliation{Dept. of Physics, TU Dortmund University, D-44221 Dortmund, Germany}
\affiliation{Dept. of Physics and Astronomy, Michigan State University, East Lansing, MI 48824, USA}
\affiliation{Dept. of Physics, University of Alberta, Edmonton, Alberta, Canada T6G 2E1}
\affiliation{Erlangen Centre for Astroparticle Physics, Friedrich-Alexander-Universit{\"a}t Erlangen-N{\"u}rnberg, D-91058 Erlangen, Germany}
\affiliation{Physik-department, Technische Universit{\"a}t M{\"u}nchen, D-85748 Garching, Germany}
\affiliation{D{\'e}partement de physique nucl{\'e}aire et corpusculaire, Universit{\'e} de Gen{\`e}ve, CH-1211 Gen{\`e}ve, Switzerland}
\affiliation{Dept. of Physics and Astronomy, University of Gent, B-9000 Gent, Belgium}
\affiliation{Dept. of Physics and Astronomy, University of California, Irvine, CA 92697, USA}
\affiliation{Karlsruhe Institute of Technology, Institute for Astroparticle Physics, D-76021 Karlsruhe, Germany }
\affiliation{Karlsruhe Institute of Technology, Institute of Experimental Particle Physics, D-76021 Karlsruhe, Germany }
\affiliation{Dept. of Physics, Engineering Physics, and Astronomy, Queen's University, Kingston, ON K7L 3N6, Canada}
\affiliation{Dept. of Physics and Astronomy, University of Kansas, Lawrence, KS 66045, USA}
\affiliation{Department of Physics and Astronomy, UCLA, Los Angeles, CA 90095, USA}
\affiliation{Centre for Cosmology, Particle Physics and Phenomenology - CP3, Universit{\'e} catholique de Louvain, Louvain-la-Neuve, Belgium}
\affiliation{Department of Physics, Mercer University, Macon, GA 31207-0001, USA}
\affiliation{Dept. of Astronomy, University of Wisconsin{\textendash}Madison, Madison, WI 53706, USA}
\affiliation{Dept. of Physics and Wisconsin IceCube Particle Astrophysics Center, University of Wisconsin{\textendash}Madison, Madison, WI 53706, USA}
\affiliation{Institute of Physics, University of Mainz, Staudinger Weg 7, D-55099 Mainz, Germany}
\affiliation{Department of Physics, Marquette University, Milwaukee, WI, 53201, USA}
\affiliation{Institut f{\"u}r Kernphysik, Westf{\"a}lische Wilhelms-Universit{\"a}t M{\"u}nster, D-48149 M{\"u}nster, Germany}
\affiliation{Bartol Research Institute and Dept. of Physics and Astronomy, University of Delaware, Newark, DE 19716, USA}
\affiliation{Dept. of Physics, Yale University, New Haven, CT 06520, USA}
\affiliation{Dept. of Physics, University of Oxford, Parks Road, Oxford OX1 3PU, UK}
\affiliation{Dept. of Physics, Drexel University, 3141 Chestnut Street, Philadelphia, PA 19104, USA}
\affiliation{Physics Department, South Dakota School of Mines and Technology, Rapid City, SD 57701, USA}
\affiliation{Dept. of Physics, University of Wisconsin, River Falls, WI 54022, USA}
\affiliation{Dept. of Physics and Astronomy, University of Rochester, Rochester, NY 14627, USA}
\affiliation{Department of Physics and Astronomy, University of Utah, Salt Lake City, UT 84112, USA}
\affiliation{Oskar Klein Centre and Dept. of Physics, Stockholm University, SE-10691 Stockholm, Sweden}
\affiliation{Dept. of Physics and Astronomy, Stony Brook University, Stony Brook, NY 11794-3800, USA}
\affiliation{Dept. of Physics, Sungkyunkwan University, Suwon 16419, Korea}
\affiliation{Institute of Basic Science, Sungkyunkwan University, Suwon 16419, Korea}
\affiliation{Dept. of Physics and Astronomy, University of Alabama, Tuscaloosa, AL 35487, USA}
\affiliation{Dept. of Astronomy and Astrophysics, Pennsylvania State University, University Park, PA 16802, USA}
\affiliation{Dept. of Physics, Pennsylvania State University, University Park, PA 16802, USA}
\affiliation{Dept. of Physics and Astronomy, Uppsala University, Box 516, S-75120 Uppsala, Sweden}
\affiliation{Dept. of Physics, University of Wuppertal, D-42119 Wuppertal, Germany}
\affiliation{DESY, D-15738 Zeuthen, Germany}

\author{R. Abbasi}
\affiliation{Department of Physics, Loyola University Chicago, Chicago, IL 60660, USA}
\author{M. Ackermann}
\affiliation{DESY, D-15738 Zeuthen, Germany}
\author{J. Adams}
\affiliation{Dept. of Physics and Astronomy, University of Canterbury, Private Bag 4800, Christchurch, New Zealand}
\author{J. A. Aguilar}
\affiliation{Universit{\'e} Libre de Bruxelles, Science Faculty CP230, B-1050 Brussels, Belgium}
\author{M. Ahlers}
\affiliation{Niels Bohr Institute, University of Copenhagen, DK-2100 Copenhagen, Denmark}
\author{M. Ahrens}
\affiliation{Oskar Klein Centre and Dept. of Physics, Stockholm University, SE-10691 Stockholm, Sweden}
\author{J.M. Alameddine}
\affiliation{Dept. of Physics, TU Dortmund University, D-44221 Dortmund, Germany}
\author{C. Alispach}
\affiliation{D{\'e}partement de physique nucl{\'e}aire et corpusculaire, Universit{\'e} de Gen{\`e}ve, CH-1211 Gen{\`e}ve, Switzerland}
\author{A. A. Alves Jr.}
\affiliation{Karlsruhe Institute of Technology, Institute for Astroparticle Physics, D-76021 Karlsruhe, Germany }
\author{N. M. Amin}
\affiliation{Bartol Research Institute and Dept. of Physics and Astronomy, University of Delaware, Newark, DE 19716, USA}
% ...

\collaboration{IceCube Collaboration}

\date{\today}

\begin{abstract}
The Sun provides an excellent target for studying spin-dependent dark matter-proton scattering due to its high matter density and abundant hydrogen content. Dark matter particles from the Galactic halo can elastically interact with Solar nuclei, resulting in their capture and thermalization in the Sun. The captured dark matter can annihilate into Standard Model particles including an observable flux of neutrinos. We present the results of a search for low-energy ($<$ 500 GeV) neutrinos correlated with the direction of the Sun using 7 years of IceCube data. This work  utilizes, for the first time, new optimized cuts to extend IceCube’s sensitivity to dark matter mass down to 5 GeV. We find no significant detection of neutrinos from the Sun. Our observations exclude capture by spin-dependent dark matter-proton scattering with cross-section down to a few times $10^{-41}$ cm$^2$, assuming there is equilibrium with annihilation into neutrinos/anti-neutrinos for dark matter masses between 5 GeV and 100 GeV. These are the strongest constraints at GeV energies for dark matter annihilation directly to neutrinos.
\end{abstract}

\maketitle

\section{Introduction}

Based on numerous observations from cosmology and astronomy, dark matter (DM) is believed to constitute over $\sim 80 \%$ of all matter in the universe \cite{planckcollaborationPlanck2018Results2020,2017IJMPD..2630012F, Bertone:2004pz,Buckley:2017ijx}. The quest to establish the particle nature of DM is also tied to observations in high energy astrophysics, including observations in neutrinos. The search for neutrinos produced by annihilations or decays of DM is one major aspect of indirect detection of DM from astrophysical objects. The Sun is particularly well-suited for such searches as it has been gravitationally capturing candidates for DM particles such as Weakly Interacting Massive Particles (WIMPs) from the surrounding halo  for its entire lifetime of ~4.5 billion years \cite{1985ApJ296679P,2009PhRvD..79j3532P,2019BAAS...51c.194N, Srednicki:1986vj,2004PhRvD..69l3505L}. These particles accumulate in the Sun, where they annihilate into standard model (SM) particles as their density builds up. This process provides a route to studying WIMP interactions with nucleons since there is time for equilibrium to be established between captures and annihilations ~\cite{Indirectdm,Garani:2017jcj,Rott:2011fh,Ajello:2011dq,2009JCAP...04..009W}. 

Given the high matter density of the Sun, the only SM particles that can escape the Sun with relatively little attenuation are neutrinos \cite{Ritz:1987mh, NG1987138, Belotsky:2008vh, Baum:2016oow,delosHeros:2015klz,Bergstrom:1998xh,Bell:2011sn}. (Secluded DM models where DM annihilation proceeds via a long-lived mediator which can decay outside the Sun into SM particles, also allow for the production of gamma rays in addition to neutrinos correlated with the direction of the Sun \cite{Allahverdi:2016fvl,Meade:2009mu, 2010PhRvD..81g5004B,2010PhRvD..81a6002S, Bell:2011sn, Feng:2016ijc, 2017PhRvD..95l3016L, Arina:2017sng, Smolinsky:2017fvb,HAWC:2018szf,Niblaeus:2019gjk, Mazziotta:2020foa}). Several experiments including Super-Kamiokande \cite{Choi:2015ara}, IceCube ~\cite{2017EPJC...77..146A,ColomBernadich:2019sqc} and ANTARES \cite{2016PhLB..759...69A, ANTARES:2016obx} have looked for neutrino signatures of DM annihilation in the Sun. These searches are especially useful for probing spin-dependent DM-proton scattering cross-sections, and have already outperformed direct detection experiments by more than an order of magnitude in terms of sensitivity. IceCube's previously published searches using three years of data already result in the world's best constraints on the spin-dependent scattering cross-section for DM mass in the range $\mathcal{O}(100)$ GeV to 10 TeV.  

Due to IceCube's optimal sensitivity to TeV--PeV neutrinos, the detector's probing of DM parameter space below 50 GeV has been limited up until now, while a large parameter space for  GeV WIMPs remains unconstrained ~\cite{Leane:2018kjk}. This work for the first time extends IceCube's reach to 5 GeV DM masses for some of the studied annihilation channels. The paper is structured as follows. Section \ref{sec:ic} describes the IceCube detector and the process of data selection used in this analysis. Section  \ref{sec:methods} presents the analysis, including the details of the signal and background estimation methods used. The results are discussed in section \ref{sec:results}. Section \ref{sec:con} presents our conclusions and places the results in context.

\section{\label{sec:ic}IceCube and DeepCore Data}
\subsection{Detector}
The IceCube Neutrino Observatory --  located at the South Pole -- consists of an array of 5160 photodetectors on 86 strings embedded within 1 km$^3$ of the Antarctic ice. Each photodetector unit -- known as a  digital optical module (DOM) -- is a downward facing photomultiplier tube (PMT) with associated electronics enclosed within a glass vessel \cite{IceCube:2010dpc}. The typical horizontal spacing between the strings is 125 m with 60 DOMs per string. The exception are the 8 strings in the bottom-center of the array known as DeepCore, which has a geometry optimized to lower the energy threshold of IceCube \cite{IceCube:2011ucd}. A higher density of high-quantum  efficiency DOMs, coupled with the outer array acting as a veto region to reject atmospheric muons makes DeepCore particularly suitable for detecting neutrinos as low as $\sim$5 GeV in energy. A detailed description of the instrumentation and signal reconstruction can be found in Refs. \cite{Aartsen_2017,Aartsen:2013vja}.        
\subsection{Event Selection}
We use IceCube and DeepCore data collected between January 1st, 2011 and January 1st, 2018 with a total live-time of 6.75 years. The event selection and reconstruction used in this analysis follows the same methods as those used in Ref. \cite{PhysRevLett.120.071801}.  The IceCube DOMs surrounding the DeepCore volume are used to veto atmospheric muons. This is achieved by rejecting events in which photons in a certain time-window are observed outside before they're detected in DeepCore. The photoelectrons detected within the DeepCore volume are fitted using a multi-dimensional likelihood to estimate the energy and direction of a neutrino event. Each event is classified as either ``track-like'' or ``cascade-like'', depending on whether the fit is better described by a $\nu_{\mu}$ charged-current (CC) interaction, or a hadronic shower with no muon resulting from neutral current interactions as well as $\nu_{\tau}/\nu_{e}$ CC interactions. An eleven variable boosted decision tree (BDT) is used to further reject atmospheric muons. 

The two main differences in the event reconstruction with respect to that in \cite{PhysRevLett.120.071801} are at the final data reduction level and are discussed here. One, we no longer require that the stopping vertex of the reconstructed muon be contained within DeepCore. Two, the boosted decision tree (BDT) cut is loosened to allow additional particles in the data sample. The purpose of the aforementioned relaxed cuts is to enhance the overall number of neutrinos in the data at the cost of an increase of $13\%$ background contamination with respect to that given in \cite{PhysRevLett.120.071801}. The final sample includes 192,212 events. This is also the first time that an IceCube analysis utilizes both ``track-like'' and ``cascade-like'' events to search for dark matter.  At the low energies considered in this work, tracks and cascades show negligible differences in their angular resolutions. The median angular resolution of events in this sample at 10 GeV is $\sim 35^\circ$ and improves to $\leq 5^\circ$ above 200 GeV.  
%%%%%%%%%%%%%%%%%%%%%%%%%%%%%%%%%%%FIGURE%%%%%%%%%%%%%%%%%%%%%%%%%%%%%%%%%%%%%%

\begin{figure*}[ht!]
\makebox[1\width][c]{
\begin{tabular}{@{}cc@{}}
\includegraphics[width=0.5\textwidth]{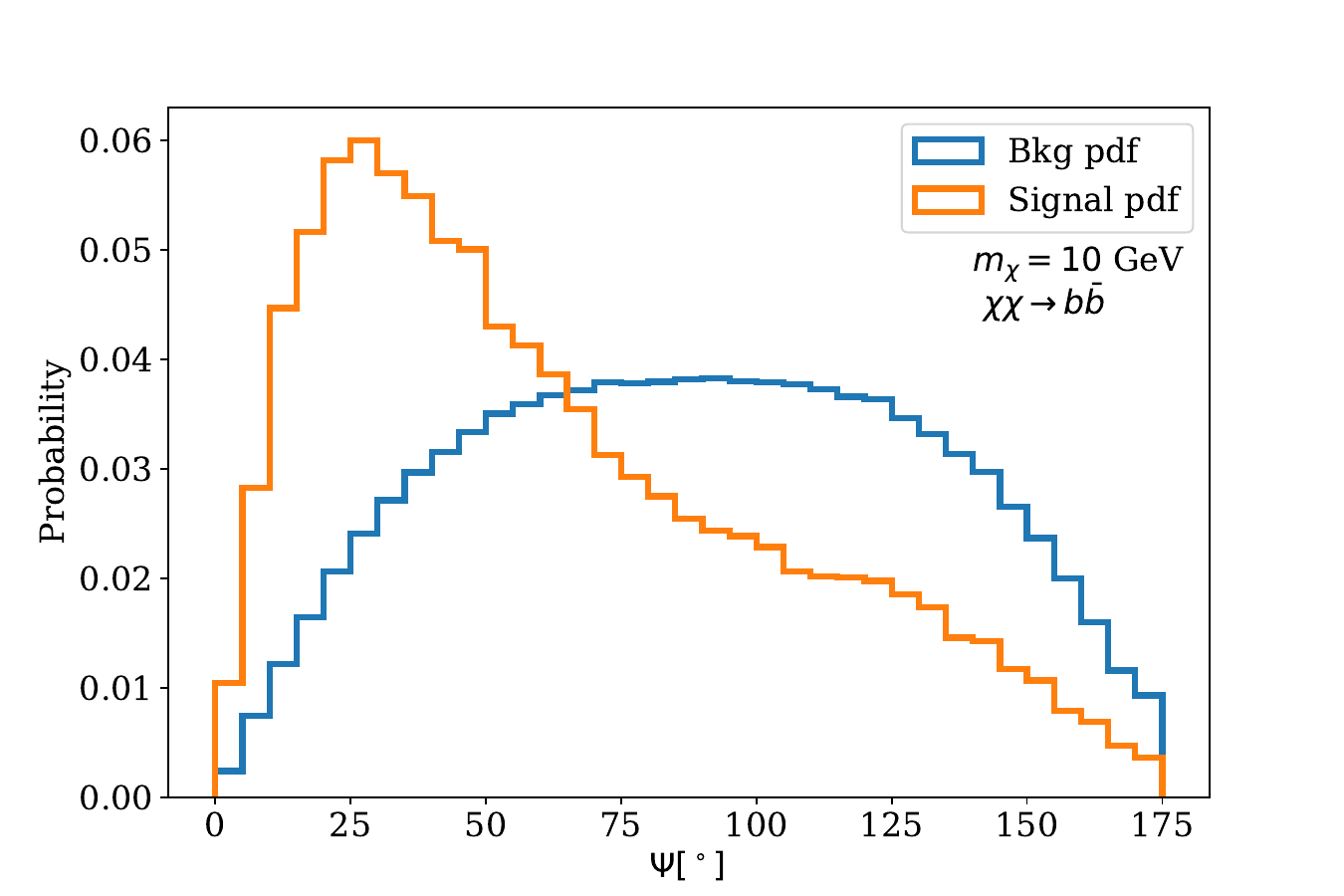} &
\includegraphics[width=0.5\textwidth]{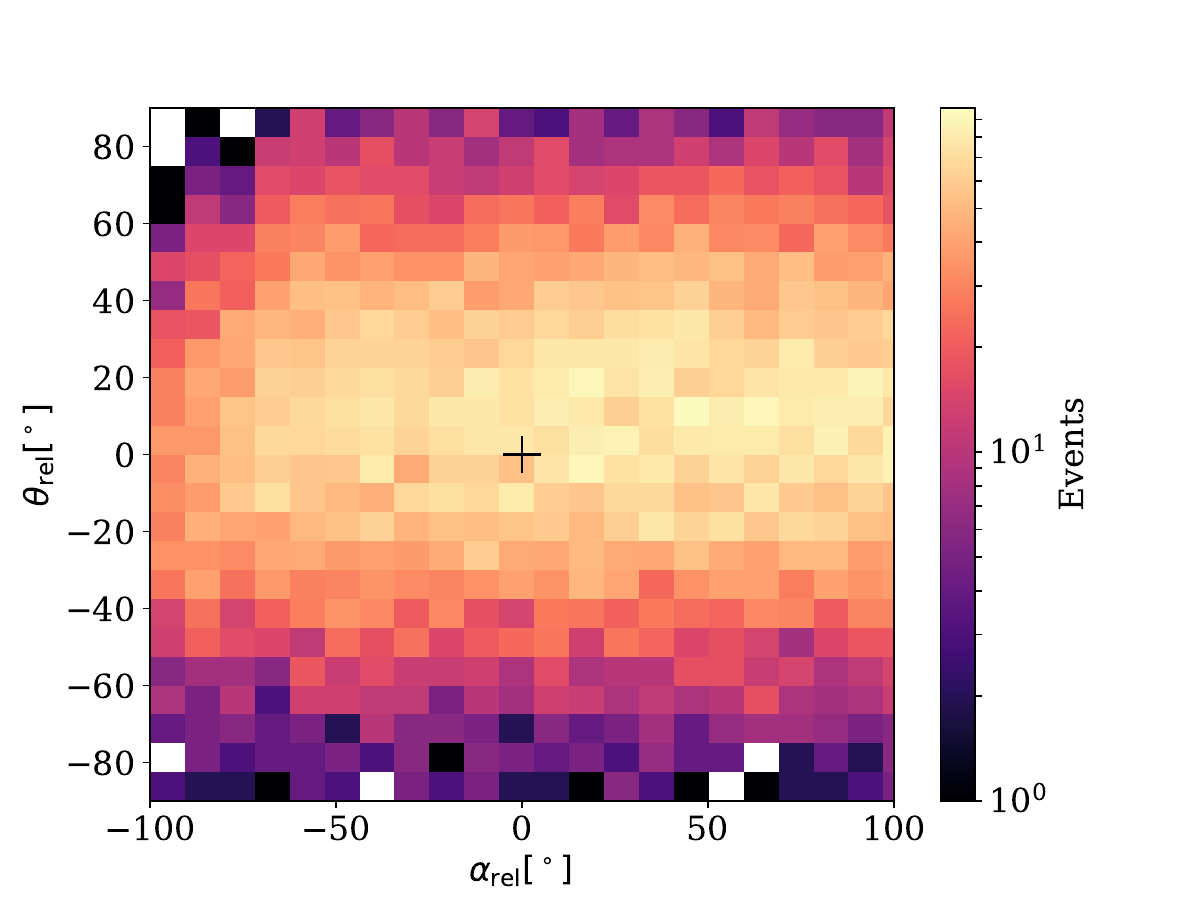} \\
\includegraphics[width=0.5\textwidth]{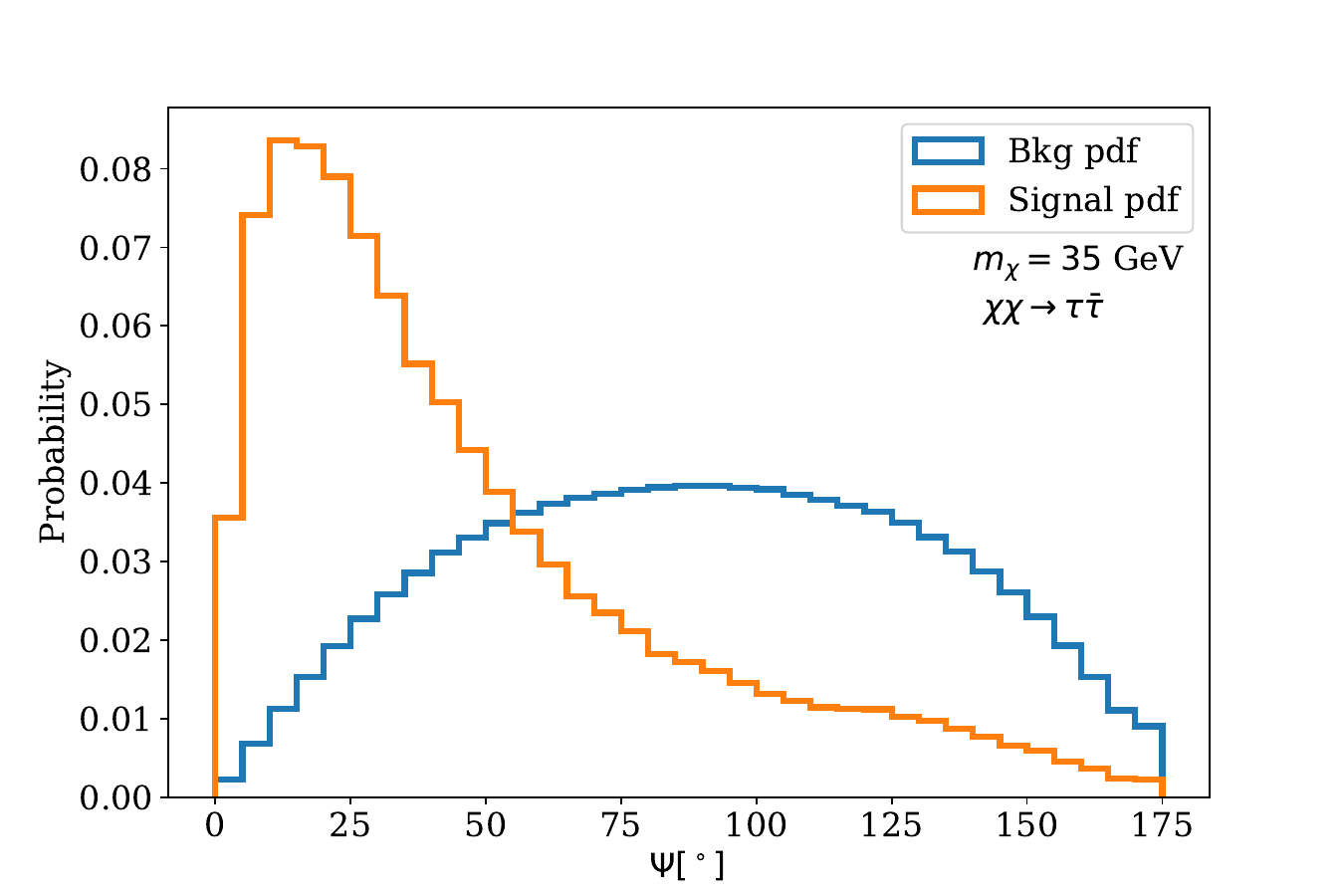} &
 \includegraphics[width=0.5\textwidth]{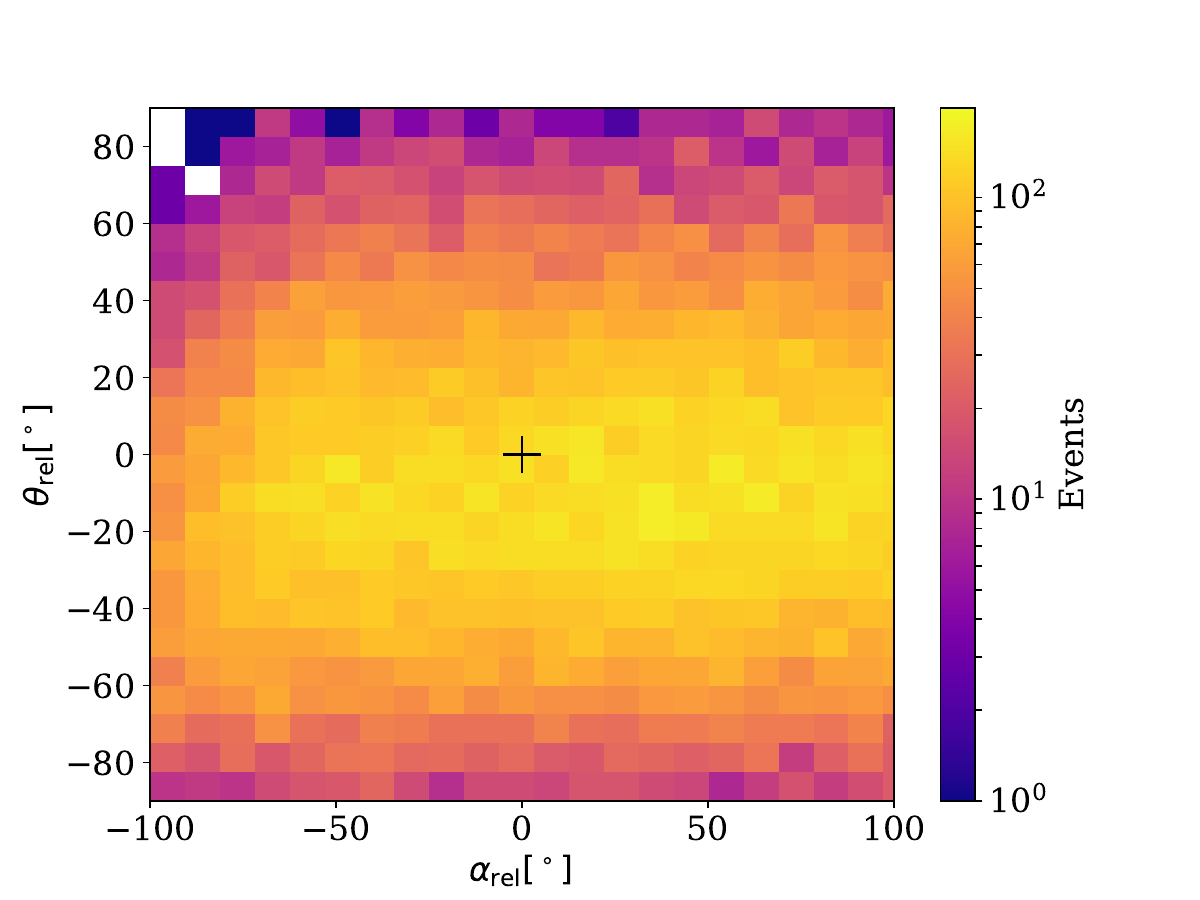}\\
\includegraphics[width=0.5\textwidth]{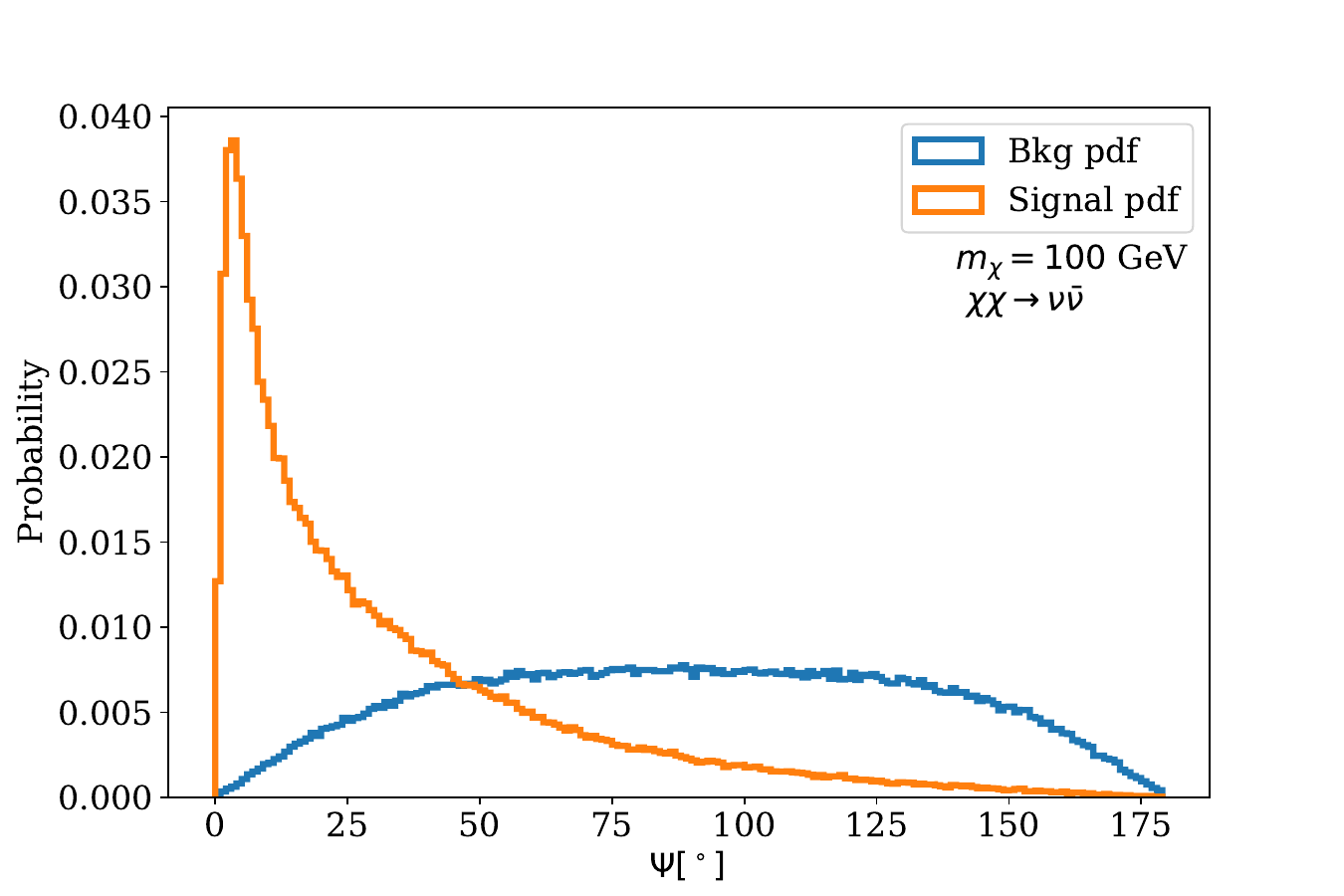} &
\includegraphics[width=0.5\textwidth]{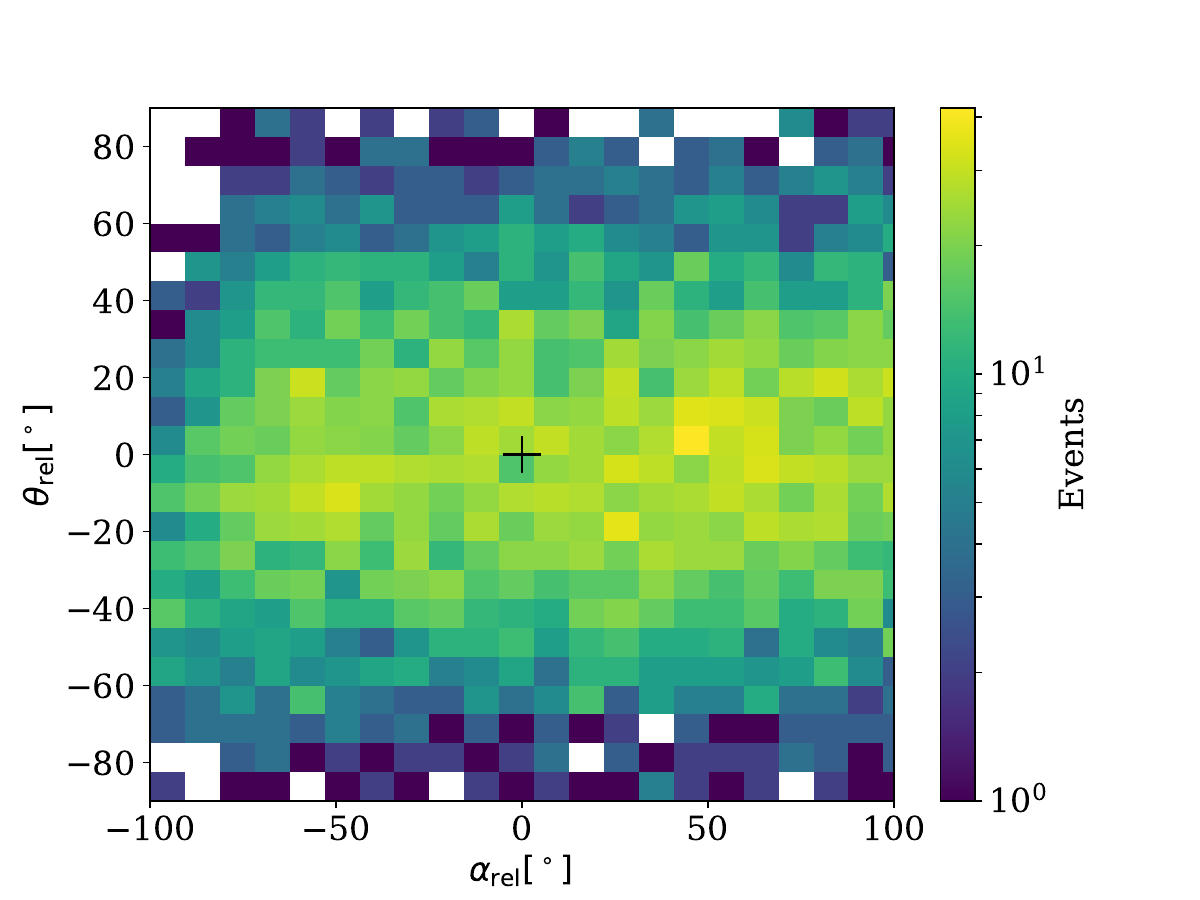} \\
\end{tabular}
}
\caption{\textbf{Left:} The PDF distributions for signal (orange) and background (blue) for three different annihilation channels and WIMP masses. The top panel corresponds to the $b\bar{b}$ annihilation channel for a 10 GeV WIMP mass, the middle panel to annihilation into $\tau^+\tau^-$ for 35 GeV WIMP mass, and the bottom panel annihilation into $\nu\bar{\nu}$ for 100 GeV WIMP mass, under the assumption of $100\%$ annihilation to the respective channel. The angle $\Psi$ represents the opening angle with respect to the Sun. \textbf{Right:} Sun-centered data maps for the corresponding channels (masses). The black cross marks the position of the Sun. $\alpha_{\rm rel}$ and $\theta_{\rm rel}$ are the azimuth and zenith angles relative to the Sun respectively.}
\label{fig:maps}
\end{figure*}

%%%%%%%%%%%%%%%%%%%%%%%%%%%%%%%%%%%FIGURE%%%%%%%%%%%%%%%%%%%%%%%%%%%%%%%%%%%%%%

%%%%%%%%%%%%%%%%%%%%%%%%%%%%%%%%%%%TABLE 1%%%%%%%%%%%%%%%%%%%%%%%%%%%%%%%%%%%%%%
\begin{table*}[ht!]
\centering
\begin{tabularx}{\textwidth}{| X | X | X | X |}
\hline
WIMP Mass (GeV) & $\tau^+\tau^-$ E$_\text{reco}$ (GeV) & $\nu\bar{\nu}$ E$_\text{reco}$  (GeV)  & $b\bar{b}$ E$_\text{reco}$ (GeV) \\
 \hline
 5 & $<$9 (7)  & 2--11 (8) & --\\
10 & 1--16 (10) & $<$23 (13)& 0--11 (8) \\
 20 &3--30  (15)& 13--39  (23) & $<$18 (11)\\
 35 & 8--50 (21)& 25--70 (38) & $<$27 (14) \\
 50 & 15--69 (29) & 42--86 (55) & 3--38 (17) \\
 100 & 30--128 (47) & 83--167 (107) & 6--70 (22)\\
 \hline
 \end{tabularx}
 \caption{The reconstructed energy ranges of neutrinos used in the search for each WIMP mass and channel. The median energy of neutrinos in each range is shown in parantheses.}
\label{tab:energies}
\end{table*}

%%%%%%%%%%%%%%%%%%%%%%%%%%%%%%%%%%%TABLE 1%%%%%%%%%%%%%%%%%%%%%%%%%%%%%%%%%%%%%%

\section{\label{sec:methods}Analysis}
We use an unbinned likelihood ratio method to search for neutrinos correlated with the direction of the Sun. The one-dimensional likelihood function is given by,

\begin{equation}
\mathcal{L} \left( n_{s}\right) = \prod_{i}^{N} \left( \frac{n_{s}}{N} S(\Psi_{i}) + \left( 1 - \frac{n_{s}}{N} \right) B(\Psi_{i}) \right),
\label{eq:1}
\end{equation}  
where $n_s$ is the number of signal neutrino events, $N$ is the total number of data events, $\Psi_i$ is the angular distance between the reconstructed direction of the $i$th event and the direction of the Sun, $S(\Psi_i)$ is the signal probability distribution function (PDF) for the $i$th data event, and $B(\Psi_i)$ is the background PDF for the $i$th data event. Given the similar angular resolutions of tracks and cascades in this sample, the likelihood does not depend on event-topology and tracks and cascades are treated identically. We also calculate a test statistic (TS), given by twice the logarithm of the ratio of the best-fit likelihood to the null (background-only) hypothesis,
\begin{equation}
\text{TS} = 2\log\frac{\mathcal{L}\left( \hat{n}_{s}\right)}{\mathcal{L} \left( n_{s} = 0 \right)},
\label{eq:2}
\end{equation}
where $\hat{n}_{s}$ is the best fit value of the number of signal events. The modeling of the signal PDF from simulation and the background PDF from randomized data are described below.

%%%%%%%%%%%%%%%%%%%%%%%%%%%%%%%%%%%TABLE 4%%%%%%%%%%%%%%%%%%%%%%%%%%%%%%%%%%%%%%
\begin{table*}[ht!]
\centering
\begin{tabularx}{\textwidth}{|l | X | X | X | X | X|}
\hline
WIMP Mass (GeV) & 10 & 20  & 35 & 50 & 100 \\
 \hline
DOM Efficiency $-6\%$ & 1.17&	1.13&	1.10&	1.09&1.03\\
DOM Efficiency $+6\%$&0.85	&0.90&	0.96&	0.95&	0.97\\
Absorption $+10\%$&1.06&	1.05	&1.03	&1.02	&0.97\\
Scattering $+10\%$&1.02&	1.06&	1.08	&1.09&	1.06\\
 \hline
 \end{tabularx}
 \caption{The ratio of sensitivity (upper limits) obtained under different systematic variations to the baseline sensitivity (upper limits) obtained in this analysis. Absolute DOM efficiency and the uncertainties in the bulk ice scattering and absorption coefficients are the most dominant systematics in this analysis.}
\label{tab:sys}
\end{table*}

%%%%%%%%%%%%%%%%%%%%%%%%%%%%%%%%%%%TABLE 4%%%%%%%%%%%%%%%%%%%%%%%%%%%%%%%%%%%%%%

\subsection{Signal and Background Probabilities}
\subsubsection{Neutrinos from DM Annihilation}
We consider only DM masses higher than ~5 GeV for which evaporation from the Sun is negligibly small \cite{1987ApJ...321..560G, Garani:2021feo}. Ignoring self-interactions, the number of DM particles in the Sun $N_{\chi}(t)$ is given by,

\begin{equation}
\frac{dN_{\chi}}{dt} = \Gamma_{\rm cap} - K_{\rm ann}N_\chi^2,
\end{equation}
where $\Gamma_{\rm cap}$ is the WIMP capture rate, and the second term expresses the annihilation rate in terms of a factor $K_{\rm ann}$, that accounts for the DM number density and the velocity-averaged annihilation cross-section ~\cite{2017JCAP...05..046W}. Once equilibrium has been reached between WIMP capture and annihilation rate, the capture rate and annihilation rate $\Gamma_{\rm ann}$ are related by,

\begin{equation}
\Gamma_{\rm cap} = 2 \Gamma_{\rm ann}.
\label{eq:cap}
\end{equation}

The factor of two accounts for the fact that every annihilation event involves two DM particles. The capture rate itself is a function of DM-proton cross-section ($\sigma_{\rm SD}$ spin-dependent and $\sigma_{\rm SI}$ spin-independent).  On the observable side, the neutrino/anti-neutrino flux at Earth from DM annihilation in the Sun $d\phi_{\nu}/dt$ is given by,

\begin{equation}
\frac{d\phi_{\nu}}{dt} = \frac{\Gamma_{\rm ann}}{4\pi D^2} \frac{dN_{\nu}}{dE},
\label{eq:dm}
\end{equation}
where $D$ is the Earth-Sun distance and $dN_{\nu}/dE$ is the spectral energy distribution of the final-state neutrinos and anti-neutrinos produced as a result of DM annihilation. This means that using the measured flux of neutrinos and the assumed DM annihilation spectra, we can constrain the annihilation rate under equilibrium (eqs. \ref{eq:cap} and \ref{eq:dm}), and therefore, the DM-proton cross-section.

We consider DM annihilation via three different channels: $b\bar{b}$, $\tau\bar{\tau}$ and $\nu\bar{\nu}$. The annihilation spectra are modeled using \textsc{\href{http://wimpsim.astroparticle.se}{WIMPSIM}}  \cite{Blennow:2007tw,Niblaeus:2019gjk}, while the neutrino interactions in the detector are simulated using \textsc{GENIE} ~\cite{Andreopoulos:2009rq}. At any given energy, we can weight the simulations by a desired flux model to calculate the total signal or background weights. The signal weight at a given energy is computed using the all-flavor neutrino spectrum from \textsc{WIMPSIM} for a given DM mass and channel, whereas the background weights are obtained from the atmospheric neutrino spectrum \cite{Honda:2006qj}.The signal PDF generation is a two-step process. First, for each annihilation channel and WIMP mass we determine an optimal range in reconstructed neutrino energy that maximizes the ratio of the summed signal weights and the square root of the background weights. Table \ref{tab:energies} lists the optimal reconstructed neutrino energy ranges for each mass and annihilation channel.  In the second step, we obtain the signal PDF by weighting the angular separation between the simulated neutrino and the reconstructed neutrino by the WIMPSIM flux at the given reconstructed neutrino energy. This procedure effectively assigns a higher weight to the neutrinos in the optimized energy range and a directional correlation with the Sun. Figure \ref{fig:maps} (left panel) illustrates the signal and background PDFs as a function of the angular separation from the Sun. 
%%%%%%%%%%%%%%%%%%%%%%%%%%%%%%%%%%%TABLE 2%%%%%%%%%%%%%%%%%%%%%%%%%%%%%%%%%%%%%%
\begin{table*}[ht!]
\begin{ruledtabular}
%\begin{tabular}{c|c|c|c|c|c|c|c|c|c}
\begin{tabular}{p{10mm}|p{16mm}|p{16mm}|p{19mm}|p{16mm}|p{16mm}|p{19mm}|p{16mm}|p{16mm}|p{19mm}}
 &\multicolumn{3}{c|}{$\mathbf{b\bar{b}}$}&\multicolumn{3}{c|}{$\boldsymbol{\tau}$ $\boldsymbol{\bar{\tau}}$} &\multicolumn{3}{c}{$\boldsymbol{\nu} $$\boldsymbol{\bar{\nu}}$}\\
 
Mass (GeV) &  $\sigma_{\rm SI}$ [cm$^{2}$] $\times 10^{-41}$ & $\sigma_{\rm SD}$ [cm$^{2}$] $\times 10^{-39}$   & $\sigma_{\rm SD}^{\rm Exp.}$  [cm$^{2}$]  $\times 10^{-39}$  &$\sigma_{\rm SI}$ [cm$^{2}$] $\times 10^{-41}$ & $\sigma_{\rm SD}$ [cm$^{2}$] $\times 10^{-39}$ & $\sigma_{\rm SD}^{\rm Exp.}$  [cm$^{2}$]  $\times 10^{-39}$ & $\sigma_{\rm SI}$ [cm$^{2}$] $\times 10^{-41}$ & $\sigma_{\rm SD}$ [cm$^{2}$]  $\times 10^{-39}$ & $\sigma_{\rm SD}^{\rm Exp.}$  [cm$^{2}$]  $\times 10^{-39}$\\ \hline

 5& - & - & - & $5.34 $ & $1.33$ & $1.38 $ & $0.38$ & $0.092$ & $0.23$\\
 
 10 &16.6 & 8.39  & 10.8  &0.29 & 0.15 & $0.21$  & 0.04 & 0.029   & $0.057 $ \\
 
 20& 1.54 & 1.57  & 2.53  &0.05 & 0.05 & $0.08$ & 0.02 &0.014 & $0.027 $\\
 
 35& 0.54 & 0.93  &  1.50  &0.02 &  0.03 & $0.05$ & 0.01 &0.012 & $0.022$ \\
 
50 & 0.34 & 0.80  & 1.29 &0.009 &  0.02 & $0.04$  & 0.004 &0.011  & $0.020$ \\

100 & 0.29  & 1.12  & 1.23  &0.008 & 0.03 & $0.04$ & 0.005 & 0.022 & $0.024$  \\

\end{tabular}
\end{ruledtabular}
\caption{90\% C.L limits on the spin-independent and spin-dependent dark matter-proton cross-section for DM annihilation to $b\bar{b}$ (left), $\tau^+\tau^-$ (center) and $\nu\bar{\nu}$. The expected sensitivity from an ensemble of background-only observations is also shown under $\sigma_{\rm SD}^{\rm Exp.}$  [cm$^{2}$] for each channel and DM mass.}
\label{tab:lim}
\end{table*}

%%%%%%%%%%%%%%%%%%%%%%%%%%%%%%%%%%%TABLE 2%%%%%%%%%%%%%%%%%%%%%%%%%%%%%%%%%%%%%%

\subsubsection{Background Estimation}
The background PDFs are parameterized as a function of the angular separation from the Sun. For every event in the data, 30 azimuth angles are randomly sampled from a uniform distribution. These 30 angles are then combined with the Sun zenith angle to generate a random ``fake'' Sun position vector. The angle between the reconstructed neutrino direction and the randomized Sun direction is then used to fill the background PDF histogram. This process ensures that for any given position of the Sun, the background is estimated by randomizing the event directions with respect to the trajectory of the Sun (Fig. ~\ref{fig:maps}).

%%%%%%%%%%%%%%%%%%%%%%%%%%%%%%%%%%%TABLE 3%%%%%%%%%%%%%%%%%%%%%%%%%%%%%%%%%%%%%%
\begin{table*}[ht!]
\begin{ruledtabular}
\begin{tabular}{c|c|c|c}
&\multicolumn{1}{c|}{$\mathbf{b\bar{b}}$}&\multicolumn{1}{c|}{$\boldsymbol{\tau}$ $\boldsymbol{\bar{\tau}}$} &\multicolumn{1}{c}{$\boldsymbol{\nu} $$\boldsymbol{\bar{\nu}}$}\\
 
Mass (GeV) &$\Gamma_{\rm ann}$ [s$^{-1}]$ $\times 10^{23}$ & $\Gamma_{\rm ann}$ [s$^{-1}$] $\times 10^{23}$ &  $\Gamma_{\rm ann}$ [s$^{-1}$]  $\times 10^{23}$ \\ \hline

 5& $139$ & $139.3$ \\
 
 10 & 396 &7.0 &  1.37   \\
 
 20& 29.7 &0.97  & 0.27  \\
 
 35& 7.41  &0.22 &  0.09  \\
 
50 & 3.51  &0.096  & 0.05  \\

100 & 1.39  &0.038 & 0.027  \\

\end{tabular}
\end{ruledtabular}
\caption{90\% C.L limits on annihilation rate for DM annihilation to $b\bar{b}$ (left), $\tau^+\tau^-$ (center) and $\nu\bar{\nu}$.}
\label{tab:lim2}
\end{table*}

%%%%%%%%%%%%%%%%%%%%%%%%%%%%%%%%%%%TABLE 3%%%%%%%%%%%%%%%%%%%%%%%%%%%%%%%%%%%%%%

%%%%%%%%%%%%%%%%%%%%%%%%%%%%%%%%%%%FIGURE%%%%%%%%%%%%%%%%%%%%%%%%%%%%%%%%%%%%%%
\begin{figure*}[ht!]
\centering
\includegraphics[width=0.99\textwidth]{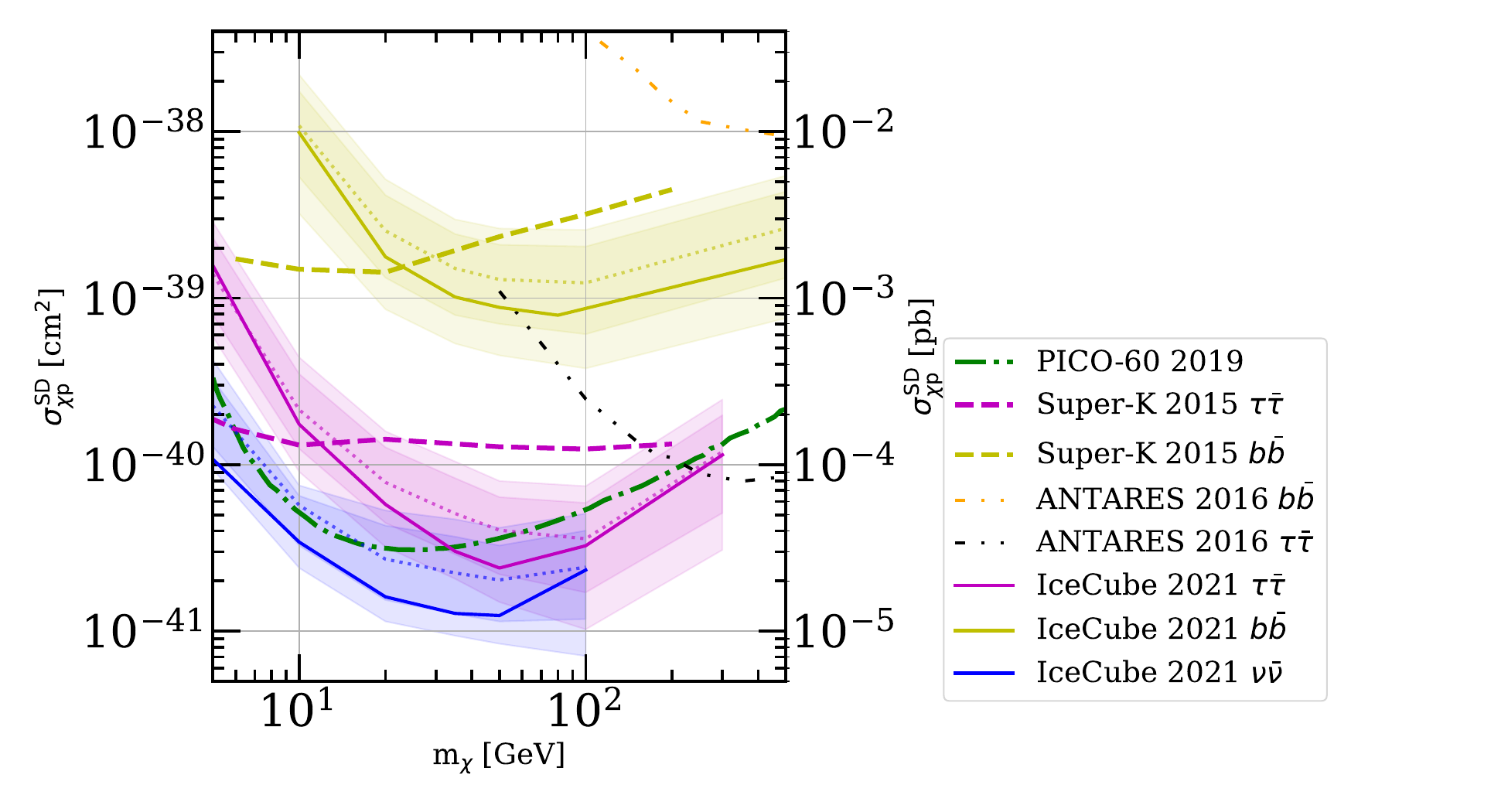}
\caption{90\% upper limits (solid lines) and expected sensitivity (dotted) on the spin-dependent cross-section as a function of WIMP mass obtained by 7 years of IceCube DeepCore data in this work. We validated the analysis up to 500 GeV and 300 GeV for $b\bar{b}$ and $\tau^+\tau^-$ but only show up to 100 GeV in the tables for consistency.The dark and light shaded bands show the central $68\%$ and $95\%$ expected limits respectively. Also shown are limits from the Super-K \cite{Choi:2015ara}, PICO-60 \cite{PhysRevD.100.022001} and ANTARES \cite{Adrian-Martinez:2016gti} experiments.} 
\label{fig:limits}
\end{figure*}

%%%%%%%%%%%%%%%%%%%%%%%%%%%%%%%%%%%FIGURE%%%%%%%%%%%%%%%%%%%%%%%%%%%%%%%%%%%%%%

\section{\label{sec:results}Results}
For all three annihilation channels, and DM masses between 5 GeV and 100 GeV (up to 500 GeV for cross-checks), we determine the best-fit number of signal event, $n_s$ that maximizes the likelihood in equation \ref{eq:1}. We obtain no statistically significant deviation from the expected background for any of the masses and channels we scanned. Figure \ref{fig:maps} (right panel) shows the observed distribution of events in a $200^\circ$ by $180^\circ$ region in Sun-centered coordinates. The highest TS obtained for any test was 0.11 for a mass of 300 GeV with DM annihilating to $\tau^+\tau^-$. We note that such an under-fluctuation of data across all tests we performed is not unlikely given that the tests are highly correlated. From background-only simulations, we expect all masses for a given channel to show a TS = 0, $5\%$ of the time.

\subsection{Systematic Uncertainties} 
The results presented in this work are sensitive to systematic uncertainties due to detector effects. The systematic uncertainties affect the overall event rate, as well as the angular and energy resolutions in the analysis. In order to study how these effects propagate into the signal PDFs and finally the upper limits on the DM-proton scattering cross-section, we repeat all the analysis steps on several simulated datasets. Each simulation was produced by varying the parameters of photon propagation at the detector, the DOM efficiency and the models of hole-ice (surrounding the strings) and the bulk ice (between the strings) up to $\pm 10 \%$. We then compare the sensitivity obtained in these simulations to that obtained from the baseline case. Table \ref{tab:sys} describes the effect on the sensitivity for each WIMP mass for the two most notable systematics, for annihilation to $b\bar{b}$ (other channels show similar trends).  At low masses (~10 GeV), the most dominant systematic -- DOM efficiency \cite{IceCube:2010dpc} -- degrades the sensitivity up to 20\%. At 100 GeV, the biggest impact is due to the modeling of bulk ice properties, such as the scattering and absorption of photons by ice \cite{Ackermann:2006pva, Fiedlschuster:2019unl}. The effect is below 8\%. 

\subsection{Constraints}
We set 90\% upper limits on $n_s$ and the annihilation rate $\Gamma_{\rm ann}$ [s$^{-1}$] of DM. The limits on annihilation rate are then converted to limits on the spin-dependent and spin-independent DM-proton cross-sections following \cite{Jungman:1995df}. Tables \ref{tab:lim} and \ref{tab:lim2} summarize these results. Figure \ref{fig:limits} shows the limits on the spin-dependent cross-section as a function of DM mass. For each mass, we show the least constraining limits as obtained under the largest systematic variation for the respective mass (Table \ref{tab:sys}). The differences between the limits for different channels depend on their spectral energy distributions relative to IceCube energy threshold.  The differences between the limits for different masses are related to IceCube's varying angular resolution with energy. In particular, poorer angular resolution ($\sim 35^\circ$)  for neutrinos below $\sim 10$ GeV, results in an increased number of background events in the search region, worsening the limits for lower masses and softer channels. For any given channel, IceCube limits on the spin-dependent WIMP-proton cross section presented in this paper are world-leading and are the strictest so far among indirect DM search experiments. IceCube is particularly sensitive to direct annihilation of DM into neutrinos and the constraints for this channel are stronger than those obtained via direct detection \cite{PhysRevD.100.022001} 

The predicted flux of solar atmospheric neutrinos is, in principle, a background for dark matter searches from the Sun \cite{Edsjo:2017kjk,2017JCAP...07..024A,2017PhRvD..96j3006N}. However, as shown in Ref. \cite{Aartsen_2021}, IceCube is not yet sensitive enough to detect the expected flux of neutrinos from cosmic ray interactions in the Sun. In fact, compared to the sensitivity required \cite{2017JCAP...07..024A,2017PhRvD..96j3006N}, the cross-section limits reported in this work are still nearly two orders of magnitude higher.

\section{\label{sec:con}Conclusion}
We present a new analysis of low-energy neutrino data from the IceCube DeepCore detector to probe spin-dependent dark matter-proton scattering and dark matter annihilation rate in the Sun. Our limits are some of the strongest in the world for a range of dark matter masses between 5 GeV and 100 GeV. The work demonstrates that neutrino telescopes even with limited statistics and angular resolution at low-energies can still provide a powerful probe of new physics. The DM limits are also a powerful probe of the coupling constants of the non-relativistic effective field theory of dark matter-nucleon interactions, including velocity- and momentum-dependent interactions \cite{Peters:2021afm}.
\begin{acknowledgments}
The IceCube Collaboration acknowledges the significant contributions to this manuscript from Garrett Neer and Mehr Un Nisa. 
USA {\textendash} U.S. National Science Foundation-Office of Polar Programs, U.S. National Science Foundation-Physics Division, U.S. National Science Foundation-EPSCoR, Wisconsin Alumni Research Foundation, Center for High Throughput Computing (CHTC) at the University of Wisconsin{\textendash}Madison, Open Science Grid (OSG), Extreme Science and Engineering Discovery Environment (XSEDE), Frontera computing project at the Texas Advanced Computing Center, U.S. Department of Energy-National Energy Research Scientific Computing Center, Particle astrophysics research computing center at the University of Maryland, Institute for Cyber-Enabled Research at Michigan State University,
and Astroparticle physics computational facility at Marquette University; Belgium {\textendash} Funds for Scientific Research (FRS-FNRS and FWO),
FWO Odysseus and Big Science programmes, and Belgian Federal Science Policy Office (Belspo), Germany {\textendash} Bundesministerium f{\"u}r Bildung und Forschung (BMBF), Deutsche Forschungsgemeinschaft (DFG), Helmholtz Alliance for Astroparticle Physics (HAP), Initiative and Networking Fund of the Helmholtz Association, Deutsches Elektronen Synchrotron (DESY), and High Performance Computing cluster of the RWTH Aachen; Sweden {\textendash} Swedish Research Council, Swedish Polar Research Secretariat, Swedish National Infrastructure for Computing (SNIC), and Knut and Alice Wallenberg Foundation; Australia {\textendash} Australian Research Council; Canada {\textendash} Natural Sciences and Engineering Research Council of Canada, Calcul Qu{\'e}bec, Compute Ontario, Canada Foundation for Innovation, WestGrid, and Compute Canada; Denmark {\textendash} Villum Fonden and Carlsberg Foundation; New Zealand {\textendash} Marsden Fund; Japan {\textendash} Japan Society for Promotion of Science (JSPS) and Institute for Global Prominent Research (IGPR) of Chiba University; Korea {\textendash} National Research Foundation of Korea (NRF); Switzerland {\textendash} Swiss National Science Foundation (SNSF); United Kingdom {\textendash} Department of Physics, University of Oxford.
\end{acknowledgments}

%\include{acks.tex}

% Create the reference section using BibTeX:
\bibliography{mbib}

%apsrev4-2.bst 2019-01-14 (MD) hand-edited version of apsrev4-1.bst
%Control: key (0)
%Control: author (8) initials jnrlst
%Control: editor formatted (1) identically to author
%Control: production of article title (0) allowed
%Control: page (0) single
%Control: year (1) truncated
%Control: production of eprint (0) enabled
\begin{thebibliography}{59}%
\makeatletter
\providecommand \@ifxundefined [1]{%
 \@ifx{#1\undefined}
}%
\providecommand \@ifnum [1]{%
 \ifnum #1\expandafter \@firstoftwo
 \else \expandafter \@secondoftwo
 \fi
}%
\providecommand \@ifx [1]{%
 \ifx #1\expandafter \@firstoftwo
 \else \expandafter \@secondoftwo
 \fi
}%
\providecommand \natexlab [1]{#1}%
\providecommand \enquote  [1]{``#1''}%
\providecommand \bibnamefont  [1]{#1}%
\providecommand \bibfnamefont [1]{#1}%
\providecommand \citenamefont [1]{#1}%
\providecommand \href@noop [0]{\@secondoftwo}%
\providecommand \href [0]{\begingroup \@sanitize@url \@href}%
\providecommand \@href[1]{\@@startlink{#1}\@@href}%
\providecommand \@@href[1]{\endgroup#1\@@endlink}%
\providecommand \@sanitize@url [0]{\catcode `\\12\catcode `\$12\catcode
  `\&12\catcode `\#12\catcode `\^12\catcode `\_12\catcode `\%12\relax}%
\providecommand \@@startlink[1]{}%
\providecommand \@@endlink[0]{}%
\providecommand \url  [0]{\begingroup\@sanitize@url \@url }%
\providecommand \@url [1]{\endgroup\@href {#1}{\urlprefix }}%
\providecommand \urlprefix  [0]{URL }%
\providecommand \Eprint [0]{\href }%
\providecommand \doibase [0]{https://doi.org/}%
\providecommand \selectlanguage [0]{\@gobble}%
\providecommand \bibinfo  [0]{\@secondoftwo}%
\providecommand \bibfield  [0]{\@secondoftwo}%
\providecommand \translation [1]{[#1]}%
\providecommand \BibitemOpen [0]{}%
\providecommand \bibitemStop [0]{}%
\providecommand \bibitemNoStop [0]{.\EOS\space}%
\providecommand \EOS [0]{\spacefactor3000\relax}%
\providecommand \BibitemShut  [1]{\csname bibitem#1\endcsname}%
\let\auto@bib@innerbib\@empty
%</preamble>
\bibitem [{\citenamefont {Aghanim}\ \emph {et~al.}(2020)\citenamefont {Aghanim}
  \emph {et~al.}}]{planckcollaborationPlanck2018Results2020}%
  \BibitemOpen
  \bibfield  {author} {\bibinfo {author} {\bibfnamefont {N.}~\bibnamefont
  {Aghanim}} \emph {et~al.} (\bibinfo {collaboration} {Planck}),\ }\bibfield
  {title} {\bibinfo {title} {Planck 2018 results. {{VI}}. {{Cosmological}}
  parameters},\ }\href {https://doi.org/10.1051/0004-6361/201833910} {\bibfield
   {journal} {\bibinfo  {journal} {A\&A}\ }\textbf {\bibinfo {volume} {641}},\
  \bibinfo {pages} {A6} (\bibinfo {year} {2020})},\ \Eprint
  {https://arxiv.org/abs/1807.06209} {arXiv:1807.06209} \BibitemShut {NoStop}%
\bibitem [{\citenamefont {Freese}(2017)}]{2017IJMPD..2630012F}%
  \BibitemOpen
  \bibfield  {author} {\bibinfo {author} {\bibfnamefont {K.}~\bibnamefont
  {Freese}},\ }\bibfield  {title} {\bibinfo {title} {Status of dark matter in
  the universe},\ }\href {https://doi.org/10.1142/S0218271817300129} {\bibfield
   {journal} {\bibinfo  {journal} {International Journal of Modern Physics D}\
  }\textbf {\bibinfo {volume} {26}},\ \bibinfo {eid} {1730012-223} (\bibinfo
  {year} {2017})},\ \Eprint {https://arxiv.org/abs/1701.01840}
  {arXiv:1701.01840} \BibitemShut {NoStop}%
\bibitem [{\citenamefont {Bertone}\ \emph {et~al.}(2005)\citenamefont
  {Bertone}, \citenamefont {Hooper},\ and\ \citenamefont
  {Silk}}]{Bertone:2004pz}%
  \BibitemOpen
  \bibfield  {author} {\bibinfo {author} {\bibfnamefont {G.}~\bibnamefont
  {Bertone}}, \bibinfo {author} {\bibfnamefont {D.}~\bibnamefont {Hooper}},\
  and\ \bibinfo {author} {\bibfnamefont {J.}~\bibnamefont {Silk}},\ }\bibfield
  {title} {\bibinfo {title} {Particle dark matter: {{Evidence}}, candidates and
  constraints},\ }\href {https://doi.org/10.1016/j.physrep.2004.08.031}
  {\bibfield  {journal} {\bibinfo  {journal} {Phys. Rept.}\ }\textbf {\bibinfo
  {volume} {405}},\ \bibinfo {pages} {279} (\bibinfo {year} {2005})},\ \Eprint
  {https://arxiv.org/abs/hep-ph/0404175} {arXiv:hep-ph/0404175} \BibitemShut
  {NoStop}%
%%CITATION = HEP-PH/0404175;%%
\bibitem [{\citenamefont {Buckley}\ and\ \citenamefont
  {Peter}(2018)}]{Buckley:2017ijx}%
  \BibitemOpen
  \bibfield  {author} {\bibinfo {author} {\bibfnamefont {M.~R.}\ \bibnamefont
  {Buckley}}\ and\ \bibinfo {author} {\bibfnamefont {A.~H.~G.}\ \bibnamefont
  {Peter}},\ }\bibfield  {title} {\bibinfo {title} {{Gravitational probes of
  dark matter physics}},\ }\href
  {https://doi.org/10.1016/j.physrep.2018.07.003} {\bibfield  {journal}
  {\bibinfo  {journal} {Phys. Rept.}\ }\textbf {\bibinfo {volume} {761}},\
  \bibinfo {pages} {1} (\bibinfo {year} {2018})},\ \Eprint
  {https://arxiv.org/abs/1712.06615} {arXiv:1712.06615 [astro-ph.CO]}
  \BibitemShut {NoStop}%
\bibitem [{\citenamefont {Press}\ and\ \citenamefont
  {Spergel}(1985)}]{1985ApJ296679P}%
  \BibitemOpen
  \bibfield  {author} {\bibinfo {author} {\bibfnamefont {W.~H.}\ \bibnamefont
  {Press}}\ and\ \bibinfo {author} {\bibfnamefont {D.~N.}\ \bibnamefont
  {Spergel}},\ }\bibfield  {title} {\bibinfo {title} {Capture by the sun of a
  galactic population of weakly interacting, massive particles},\ }\href
  {https://doi.org/10.1086/163485} {\bibfield  {journal} {\bibinfo  {journal}
  {\apj}\ }\textbf {\bibinfo {volume} {296}},\ \bibinfo {pages} {679} (\bibinfo
  {year} {1985})}\BibitemShut {NoStop}%
\bibitem [{\citenamefont {Peter}(2009)}]{2009PhRvD..79j3532P}%
  \BibitemOpen
  \bibfield  {author} {\bibinfo {author} {\bibfnamefont {A.~H.~G.}\
  \bibnamefont {Peter}},\ }\bibfield  {title} {\bibinfo {title} {Dark matter in
  the {{Solar System}}. {{II}}. {{WIMP}} annihilation rates in the {{Sun}}},\
  }\href {https://doi.org/10.1103/PhysRevD.79.103532} {\bibfield  {journal}
  {\bibinfo  {journal} {Phys. Rev. D}\ }\textbf {\bibinfo {volume} {79}},\
  \bibinfo {eid} {103532} (\bibinfo {year} {2009})},\ \Eprint
  {https://arxiv.org/abs/0902.1347} {arXiv:0902.1347} \BibitemShut {NoStop}%
\bibitem [{\citenamefont {{Nisa}}\ \emph {et~al.}(2019)\citenamefont {{Nisa}},
  \citenamefont {{Beacom}}, \citenamefont {{Peter}}, \citenamefont {{Leane}},
  \citenamefont {{Linden}}, \citenamefont {{Ng}},\ and\ \citenamefont
  {{Zhou}}}]{2019BAAS...51c.194N}%
  \BibitemOpen
  \bibfield  {author} {\bibinfo {author} {\bibfnamefont {M.}~\bibnamefont
  {{Nisa}}}, \bibinfo {author} {\bibfnamefont {J.}~\bibnamefont {{Beacom}}},
  \bibinfo {author} {\bibfnamefont {A.}~\bibnamefont {{Peter}}}, \bibinfo
  {author} {\bibfnamefont {R.}~\bibnamefont {{Leane}}}, \bibinfo {author}
  {\bibfnamefont {T.}~\bibnamefont {{Linden}}}, \bibinfo {author}
  {\bibfnamefont {K.}~\bibnamefont {{Ng}}},\ and\ \bibinfo {author}
  {\bibfnamefont {B.}~\bibnamefont {{Zhou}}},\ }\bibfield  {title} {\bibinfo
  {title} {{The Sun at GeV-TeV Energies: A New Laboratory for Astroparticle
  Physics}},\ }\href@noop {} {\bibfield  {journal} {\bibinfo  {journal} {The
  Bulletin of the American Astronomical Society}\ }\textbf {\bibinfo {volume}
  {51}},\ \bibinfo {eid} {194} (\bibinfo {year} {2019})},\ \Eprint
  {https://arxiv.org/abs/1903.06349} {arXiv:1903.06349 [astro-ph.HE]}
  \BibitemShut {NoStop}%
\bibitem [{\citenamefont {Srednicki}\ \emph {et~al.}(1987)\citenamefont
  {Srednicki}, \citenamefont {Olive},\ and\ \citenamefont
  {Silk}}]{Srednicki:1986vj}%
  \BibitemOpen
  \bibfield  {author} {\bibinfo {author} {\bibfnamefont {M.}~\bibnamefont
  {Srednicki}}, \bibinfo {author} {\bibfnamefont {K.~A.}\ \bibnamefont
  {Olive}},\ and\ \bibinfo {author} {\bibfnamefont {J.}~\bibnamefont {Silk}},\
  }\bibfield  {title} {\bibinfo {title} {{High-Energy Neutrinos from the Sun
  and Cold Dark Matter}},\ }\href
  {https://doi.org/10.1016/0550-3213(87)90020-4} {\bibfield  {journal}
  {\bibinfo  {journal} {Nucl. Phys. B}\ }\textbf {\bibinfo {volume} {279}},\
  \bibinfo {pages} {804} (\bibinfo {year} {1987})}\BibitemShut {NoStop}%
\bibitem [{\citenamefont {Lundberg}\ and\ \citenamefont
  {Edsj{\"o}}(2004)}]{2004PhRvD..69l3505L}%
  \BibitemOpen
  \bibfield  {author} {\bibinfo {author} {\bibfnamefont {J.}~\bibnamefont
  {Lundberg}}\ and\ \bibinfo {author} {\bibfnamefont {J.}~\bibnamefont
  {Edsj{\"o}}},\ }\bibfield  {title} {\bibinfo {title} {Weakly interacting
  massive particle diffusion in the solar system including solar depletion and
  its effect on {{Earth}} capture rates},\ }\href
  {https://doi.org/10.1103/PhysRevD.69.123505} {\bibfield  {journal} {\bibinfo
  {journal} {Phys. Rev. D}\ }\textbf {\bibinfo {volume} {69}},\ \bibinfo {eid}
  {123505} (\bibinfo {year} {2004})},\ \Eprint
  {https://arxiv.org/abs/astro-ph/0401113} {astro-ph/0401113} \BibitemShut
  {NoStop}%
\bibitem [{\citenamefont {Conrad}\ and\ \citenamefont
  {Reimer}(2017)}]{Indirectdm}%
  \BibitemOpen
  \bibfield  {author} {\bibinfo {author} {\bibfnamefont {J.}~\bibnamefont
  {Conrad}}\ and\ \bibinfo {author} {\bibfnamefont {O.}~\bibnamefont
  {Reimer}},\ }\bibfield  {title} {\bibinfo {title} {Indirect dark matter
  searches in gamma and cosmic rays},\ }\href
  {http://dx.doi.org/10.1038/nphys4049} {\bibfield  {journal} {\bibinfo
  {journal} {Nat. Phys.}\ }\textbf {\bibinfo {volume} {13}},\ \bibinfo {pages}
  {224} (\bibinfo {year} {2017})}\BibitemShut {NoStop}%
\bibitem [{\citenamefont {Garani}\ and\ \citenamefont
  {{Palomares-Ruiz}}(2017)}]{Garani:2017jcj}%
  \BibitemOpen
  \bibfield  {author} {\bibinfo {author} {\bibfnamefont {R.}~\bibnamefont
  {Garani}}\ and\ \bibinfo {author} {\bibfnamefont {S.}~\bibnamefont
  {{Palomares-Ruiz}}},\ }\bibfield  {title} {\bibinfo {title} {Dark matter in
  the {{Sun}}: Scattering off electrons vs nucleons},\ }\href
  {https://doi.org/10.1088/1475-7516/2017/05/007} {\bibfield  {journal}
  {\bibinfo  {journal} {JCAP}\ }\textbf {\bibinfo {volume} {1705}}\bibfield
  {number} {\bibinfo  {number} { (05)},\ \bibinfo {pages} {007}},\ }\Eprint
  {https://arxiv.org/abs/1702.02768} {arXiv:1702.02768} \BibitemShut {NoStop}%
%%CITATION = ARXIV:1702.02768;%%
\bibitem [{\citenamefont {Rott}\ \emph {et~al.}(2011)\citenamefont {Rott},
  \citenamefont {Tanaka},\ and\ \citenamefont {Itow}}]{Rott:2011fh}%
  \BibitemOpen
  \bibfield  {author} {\bibinfo {author} {\bibfnamefont {C.}~\bibnamefont
  {Rott}}, \bibinfo {author} {\bibfnamefont {T.}~\bibnamefont {Tanaka}},\ and\
  \bibinfo {author} {\bibfnamefont {Y.}~\bibnamefont {Itow}},\ }\bibfield
  {title} {\bibinfo {title} {Enhanced {{Sensitivity}} to {{Dark Matter
  Self}}-annihilations in the {{Sun}} using {{Neutrino Spectral
  Information}}},\ }\href {https://doi.org/10.1088/1475-7516/2011/09/029}
  {\bibfield  {journal} {\bibinfo  {journal} {JCAP}\ }\textbf {\bibinfo
  {volume} {1109}},\ \bibinfo {pages} {029}},\ \Eprint
  {https://arxiv.org/abs/1107.3182} {arXiv:1107.3182} \BibitemShut {NoStop}%
%%CITATION = ARXIV:1107.3182;%%
\bibitem [{\citenamefont {Ajello}\ \emph {et~al.}(2011)\citenamefont {Ajello}
  \emph {et~al.}}]{Ajello:2011dq}%
  \BibitemOpen
  \bibfield  {author} {\bibinfo {author} {\bibfnamefont {M.}~\bibnamefont
  {Ajello}} \emph {et~al.} (\bibinfo {collaboration} {Fermi LAT}),\ }\bibfield
  {title} {\bibinfo {title} {Constraints on dark matter models from a {{Fermi
  LAT}} search for high-energy cosmic-ray electrons from the {{Sun}}},\ }\href
  {https://doi.org/10.1103/PhysRevD.84.032007} {\bibfield  {journal} {\bibinfo
  {journal} {Phys. Rev.}\ }\textbf {\bibinfo {volume} {D84}},\ \bibinfo {pages}
  {032007} (\bibinfo {year} {2011})},\ \Eprint
  {https://arxiv.org/abs/1107.4272} {arXiv:1107.4272} \BibitemShut {NoStop}%
%%CITATION = ARXIV:1107.4272;%%
\bibitem [{\citenamefont {{Wikstr{\"o}m}}\ and\ \citenamefont
  {{Edsj{\"o}}}(2009)}]{2009JCAP...04..009W}%
  \BibitemOpen
  \bibfield  {author} {\bibinfo {author} {\bibfnamefont {G.}~\bibnamefont
  {{Wikstr{\"o}m}}}\ and\ \bibinfo {author} {\bibfnamefont {J.}~\bibnamefont
  {{Edsj{\"o}}}},\ }\bibfield  {title} {\bibinfo {title} {{Limits on the
  WIMP-nucleon scattering cross-section from neutrino telescopes}},\ }\href
  {https://doi.org/10.1088/1475-7516/2009/04/009} {\bibfield  {journal}
  {\bibinfo  {journal} {JCAP}\ }\textbf {\bibinfo {volume} {2009}}\bibfield
  {number} {\bibinfo  {number} { (4)},\ \bibinfo {eid} {009}},\ }\Eprint
  {https://arxiv.org/abs/0903.2986} {arXiv:0903.2986 [astro-ph.CO]}
  \BibitemShut {NoStop}%
\bibitem [{\citenamefont {Ritz}\ and\ \citenamefont
  {Seckel}(1988)}]{Ritz:1987mh}%
  \BibitemOpen
  \bibfield  {author} {\bibinfo {author} {\bibfnamefont {S.}~\bibnamefont
  {Ritz}}\ and\ \bibinfo {author} {\bibfnamefont {D.}~\bibnamefont {Seckel}},\
  }\bibfield  {title} {\bibinfo {title} {{Detailed Neutrino Spectra From Cold
  Dark Matter Annihilations in the Sun}},\ }\href
  {https://doi.org/10.1016/0550-3213(88)90660-8} {\bibfield  {journal}
  {\bibinfo  {journal} {Nucl. Phys. B}\ }\textbf {\bibinfo {volume} {304}},\
  \bibinfo {pages} {877} (\bibinfo {year} {1988})}\BibitemShut {NoStop}%
\bibitem [{\citenamefont {Ng}\ \emph {et~al.}(1987)\citenamefont {Ng},
  \citenamefont {Olive},\ and\ \citenamefont {Srednicki}}]{NG1987138}%
  \BibitemOpen
  \bibfield  {author} {\bibinfo {author} {\bibfnamefont {K.-W.}\ \bibnamefont
  {Ng}}, \bibinfo {author} {\bibfnamefont {K.~A.}\ \bibnamefont {Olive}},\ and\
  \bibinfo {author} {\bibfnamefont {M.}~\bibnamefont {Srednicki}},\ }\bibfield
  {title} {\bibinfo {title} {Dark matter induced neutrinos from the sun: Theory
  versus experiment},\ }\href
  {https://doi.org/https://doi.org/10.1016/0370-2693(87)90720-9} {\bibfield
  {journal} {\bibinfo  {journal} {Physics Letters B}\ }\textbf {\bibinfo
  {volume} {188}},\ \bibinfo {pages} {138} (\bibinfo {year}
  {1987})}\BibitemShut {NoStop}%
\bibitem [{\citenamefont {Belotsky}\ \emph {et~al.}(2009)\citenamefont
  {Belotsky}, \citenamefont {Khlopov},\ and\ \citenamefont
  {Kouvaris}}]{Belotsky:2008vh}%
  \BibitemOpen
  \bibfield  {author} {\bibinfo {author} {\bibfnamefont {K.}~\bibnamefont
  {Belotsky}}, \bibinfo {author} {\bibfnamefont {M.}~\bibnamefont {Khlopov}},\
  and\ \bibinfo {author} {\bibfnamefont {C.}~\bibnamefont {Kouvaris}},\
  }\bibfield  {title} {\bibinfo {title} {{Muon flux limits for Majorana dark
  matter from strong coupling theories}},\ }\href
  {https://doi.org/10.1103/PhysRevD.79.083520} {\bibfield  {journal} {\bibinfo
  {journal} {Phys. Rev. D}\ }\textbf {\bibinfo {volume} {79}},\ \bibinfo
  {pages} {083520} (\bibinfo {year} {2009})},\ \Eprint
  {https://arxiv.org/abs/0810.2022} {arXiv:0810.2022 [astro-ph]} \BibitemShut
  {NoStop}%
\bibitem [{\citenamefont {Baum}\ \emph {et~al.}(2017)\citenamefont {Baum},
  \citenamefont {Visinelli}, \citenamefont {Freese},\ and\ \citenamefont
  {Stengel}}]{Baum:2016oow}%
  \BibitemOpen
  \bibfield  {author} {\bibinfo {author} {\bibfnamefont {S.}~\bibnamefont
  {Baum}}, \bibinfo {author} {\bibfnamefont {L.}~\bibnamefont {Visinelli}},
  \bibinfo {author} {\bibfnamefont {K.}~\bibnamefont {Freese}},\ and\ \bibinfo
  {author} {\bibfnamefont {P.}~\bibnamefont {Stengel}},\ }\bibfield  {title}
  {\bibinfo {title} {Dark matter capture, subdominant {{WIMPs}}, and neutrino
  observatories},\ }\href {https://doi.org/10.1103/PhysRevD.95.043007}
  {\bibfield  {journal} {\bibinfo  {journal} {Phys. Rev.}\ }\textbf {\bibinfo
  {volume} {D95}},\ \bibinfo {pages} {043007} (\bibinfo {year} {2017})},\
  \Eprint {https://arxiv.org/abs/1611.09665} {arXiv:1611.09665} \BibitemShut
  {NoStop}%
%%CITATION = ARXIV:1611.09665;%%
\bibitem [{\citenamefont {de~los Heros}(2017)}]{delosHeros:2015klz}%
  \BibitemOpen
  \bibfield  {author} {\bibinfo {author} {\bibfnamefont {C.~P.}\ \bibnamefont
  {de~los Heros}},\ }\bibinfo {title} {{The Quest for Dark Matter with Neutrino
  Telescopes}}\ (\bibinfo {year} {2017})\ \Eprint
  {https://arxiv.org/abs/1511.03500} {arXiv:1511.03500 [astro-ph.HE]}
  \BibitemShut {NoStop}%
\bibitem [{\citenamefont {Bergstrom}\ \emph {et~al.}(1998)\citenamefont
  {Bergstrom}, \citenamefont {Edsjo},\ and\ \citenamefont
  {Gondolo}}]{Bergstrom:1998xh}%
  \BibitemOpen
  \bibfield  {author} {\bibinfo {author} {\bibfnamefont {L.}~\bibnamefont
  {Bergstrom}}, \bibinfo {author} {\bibfnamefont {J.}~\bibnamefont {Edsjo}},\
  and\ \bibinfo {author} {\bibfnamefont {P.}~\bibnamefont {Gondolo}},\
  }\bibfield  {title} {\bibinfo {title} {{Indirect detection of dark matter in
  km size neutrino telescopes}},\ }\href
  {https://doi.org/10.1103/PhysRevD.58.103519} {\bibfield  {journal} {\bibinfo
  {journal} {Phys. Rev. D}\ }\textbf {\bibinfo {volume} {58}},\ \bibinfo
  {pages} {103519} (\bibinfo {year} {1998})},\ \Eprint
  {https://arxiv.org/abs/hep-ph/9806293} {arXiv:hep-ph/9806293} \BibitemShut
  {NoStop}%
\bibitem [{\citenamefont {Bell}\ and\ \citenamefont
  {Petraki}(2011)}]{Bell:2011sn}%
  \BibitemOpen
  \bibfield  {author} {\bibinfo {author} {\bibfnamefont {N.~F.}\ \bibnamefont
  {Bell}}\ and\ \bibinfo {author} {\bibfnamefont {K.}~\bibnamefont {Petraki}},\
  }\bibfield  {title} {\bibinfo {title} {Enhanced neutrino signals from dark
  matter annihilation in the {{Sun}} via metastable mediators},\ }\href
  {https://doi.org/10.1088/1475-7516/2011/04/003} {\bibfield  {journal}
  {\bibinfo  {journal} {JCAP}\ }\textbf {\bibinfo {volume} {1104}},\ \bibinfo
  {pages} {003}},\ \Eprint {https://arxiv.org/abs/1102.2958} {arXiv:1102.2958}
  \BibitemShut {NoStop}%
%%CITATION = ARXIV:1102.2958;%%
\bibitem [{\citenamefont {Allahverdi}\ \emph {et~al.}(2017)\citenamefont
  {Allahverdi}, \citenamefont {Gao}, \citenamefont {Knockel},\ and\
  \citenamefont {Shalgar}}]{Allahverdi:2016fvl}%
  \BibitemOpen
  \bibfield  {author} {\bibinfo {author} {\bibfnamefont {R.}~\bibnamefont
  {Allahverdi}}, \bibinfo {author} {\bibfnamefont {Y.}~\bibnamefont {Gao}},
  \bibinfo {author} {\bibfnamefont {B.}~\bibnamefont {Knockel}},\ and\ \bibinfo
  {author} {\bibfnamefont {S.}~\bibnamefont {Shalgar}},\ }\bibfield  {title}
  {\bibinfo {title} {Indirect signals from solar dark matter annihilation to
  long-lived right-handed neutrinos},\ }\href
  {https://doi.org/10.1103/PhysRevD.95.075001} {\bibfield  {journal} {\bibinfo
  {journal} {Phys. Rev. D}\ }\textbf {\bibinfo {volume} {95}},\ \bibinfo
  {pages} {075001} (\bibinfo {year} {2017})}\BibitemShut {NoStop}%
\bibitem [{\citenamefont {Meade}\ \emph {et~al.}(2010)\citenamefont {Meade},
  \citenamefont {Nussinov}, \citenamefont {Papucci},\ and\ \citenamefont
  {Volansky}}]{Meade:2009mu}%
  \BibitemOpen
  \bibfield  {author} {\bibinfo {author} {\bibfnamefont {P.}~\bibnamefont
  {Meade}}, \bibinfo {author} {\bibfnamefont {S.}~\bibnamefont {Nussinov}},
  \bibinfo {author} {\bibfnamefont {M.}~\bibnamefont {Papucci}},\ and\ \bibinfo
  {author} {\bibfnamefont {T.}~\bibnamefont {Volansky}},\ }\bibfield  {title}
  {\bibinfo {title} {Searches for {{Long Lived Neutral Particles}}},\ }\href
  {https://doi.org/10.1007/JHEP06(2010)029} {\bibfield  {journal} {\bibinfo
  {journal} {JHEP}\ }\textbf {\bibinfo {volume} {06}},\ \bibinfo {pages}
  {029}},\ \Eprint {https://arxiv.org/abs/0910.4160} {arXiv:0910.4160}
  \BibitemShut {NoStop}%
%%CITATION = ARXIV:0910.4160;%%
\bibitem [{\citenamefont {Batell}\ \emph {et~al.}(2010)\citenamefont {Batell},
  \citenamefont {Pospelov}, \citenamefont {Ritz},\ and\ \citenamefont
  {Shang}}]{2010PhRvD..81g5004B}%
  \BibitemOpen
  \bibfield  {author} {\bibinfo {author} {\bibfnamefont {B.}~\bibnamefont
  {Batell}}, \bibinfo {author} {\bibfnamefont {M.}~\bibnamefont {Pospelov}},
  \bibinfo {author} {\bibfnamefont {A.}~\bibnamefont {Ritz}},\ and\ \bibinfo
  {author} {\bibfnamefont {Y.}~\bibnamefont {Shang}},\ }\bibfield  {title}
  {\bibinfo {title} {Solar gamma rays powered by secluded dark matter},\ }\href
  {https://doi.org/10.1103/PhysRevD.81.075004} {\bibfield  {journal} {\bibinfo
  {journal} {Phys. Rev. D}\ }\textbf {\bibinfo {volume} {81}},\ \bibinfo {eid}
  {075004} (\bibinfo {year} {2010})},\ \Eprint
  {https://arxiv.org/abs/0910.1567} {arXiv:0910.1567} \BibitemShut {NoStop}%
\bibitem [{\citenamefont {Schuster}\ \emph {et~al.}(2010)\citenamefont
  {Schuster}, \citenamefont {Toro},\ and\ \citenamefont
  {Yavin}}]{2010PhRvD..81a6002S}%
  \BibitemOpen
  \bibfield  {author} {\bibinfo {author} {\bibfnamefont {P.}~\bibnamefont
  {Schuster}}, \bibinfo {author} {\bibfnamefont {N.}~\bibnamefont {Toro}},\
  and\ \bibinfo {author} {\bibfnamefont {I.}~\bibnamefont {Yavin}},\ }\bibfield
   {title} {\bibinfo {title} {Terrestrial and solar limits on long-lived
  particles in a dark sector},\ }\href
  {https://doi.org/10.1103/PhysRevD.81.016002} {\bibfield  {journal} {\bibinfo
  {journal} {Phys. Rev. D}\ }\textbf {\bibinfo {volume} {81}},\ \bibinfo {eid}
  {016002} (\bibinfo {year} {2010})},\ \Eprint
  {https://arxiv.org/abs/0910.1602} {arXiv:0910.1602} \BibitemShut {NoStop}%
\bibitem [{\citenamefont {Feng}\ \emph {et~al.}(2016)\citenamefont {Feng},
  \citenamefont {Smolinsky},\ and\ \citenamefont {Tanedo}}]{Feng:2016ijc}%
  \BibitemOpen
  \bibfield  {author} {\bibinfo {author} {\bibfnamefont {J.~L.}\ \bibnamefont
  {Feng}}, \bibinfo {author} {\bibfnamefont {J.}~\bibnamefont {Smolinsky}},\
  and\ \bibinfo {author} {\bibfnamefont {P.}~\bibnamefont {Tanedo}},\
  }\bibfield  {title} {\bibinfo {title} {Detecting dark matter through dark
  photons from the {{Sun}}: {{Charged}} particle signatures},\ }\href
  {https://doi.org/10.1103/PhysRevD.93.115036, 10.1103/PhysRevD.96.099903}
  {\bibfield  {journal} {\bibinfo  {journal} {Phys. Rev.}\ }\textbf {\bibinfo
  {volume} {D93}},\ \bibinfo {pages} {115036} (\bibinfo {year} {2016})},\
  \Eprint {https://arxiv.org/abs/1602.01465} {arXiv:1602.01465} \BibitemShut
  {NoStop}%
%%CITATION = ARXIV:1602.01465;%%
\bibitem [{\citenamefont {Leane}\ \emph {et~al.}(2017)\citenamefont {Leane},
  \citenamefont {Ng},\ and\ \citenamefont {Beacom}}]{2017PhRvD..95l3016L}%
  \BibitemOpen
  \bibfield  {author} {\bibinfo {author} {\bibfnamefont {R.~K.}\ \bibnamefont
  {Leane}}, \bibinfo {author} {\bibfnamefont {K.~C.~Y.}\ \bibnamefont {Ng}},\
  and\ \bibinfo {author} {\bibfnamefont {J.~F.}\ \bibnamefont {Beacom}},\
  }\bibfield  {title} {\bibinfo {title} {Powerful solar signatures of
  long-lived dark mediators},\ }\href
  {https://doi.org/10.1103/PhysRevD.95.123016} {\bibfield  {journal} {\bibinfo
  {journal} {Phys. Rev. D}\ }\textbf {\bibinfo {volume} {95}},\ \bibinfo {eid}
  {123016} (\bibinfo {year} {2017})},\ \Eprint
  {https://arxiv.org/abs/1703.04629} {arXiv:1703.04629} \BibitemShut {NoStop}%
\bibitem [{\citenamefont {Arina}\ \emph {et~al.}(2017)\citenamefont {Arina},
  \citenamefont {Backovi{\'c}}, \citenamefont {Heisig},\ and\ \citenamefont
  {Lucente}}]{Arina:2017sng}%
  \BibitemOpen
  \bibfield  {author} {\bibinfo {author} {\bibfnamefont {C.}~\bibnamefont
  {Arina}}, \bibinfo {author} {\bibfnamefont {M.}~\bibnamefont {Backovi{\'c}}},
  \bibinfo {author} {\bibfnamefont {J.}~\bibnamefont {Heisig}},\ and\ \bibinfo
  {author} {\bibfnamefont {M.}~\bibnamefont {Lucente}},\ }\bibfield  {title}
  {\bibinfo {title} {Solar {$\gamma$} rays as a complementary probe of dark
  matter},\ }\href {https://doi.org/10.1103/PhysRevD.96.063010} {\bibfield
  {journal} {\bibinfo  {journal} {Phys. Rev.}\ }\textbf {\bibinfo {volume}
  {D96}},\ \bibinfo {pages} {063010} (\bibinfo {year} {2017})},\ \Eprint
  {https://arxiv.org/abs/1703.08087} {arXiv:1703.08087} \BibitemShut {NoStop}%
%%CITATION = ARXIV:1703.08087;%%
\bibitem [{\citenamefont {Smolinsky}\ and\ \citenamefont
  {Tanedo}(2017)}]{Smolinsky:2017fvb}%
  \BibitemOpen
  \bibfield  {author} {\bibinfo {author} {\bibfnamefont {J.}~\bibnamefont
  {Smolinsky}}\ and\ \bibinfo {author} {\bibfnamefont {P.}~\bibnamefont
  {Tanedo}},\ }\bibfield  {title} {\bibinfo {title} {Dark photons from captured
  inelastic dark matter annihilation: {{Charged}} particle signatures},\ }\href
  {https://doi.org/10.1103/PhysRevD.95.075015} {\bibfield  {journal} {\bibinfo
  {journal} {Phys. Rev. D}\ }\textbf {\bibinfo {volume} {95}},\ \bibinfo {eid}
  {075015} (\bibinfo {year} {2017})},\ \Eprint
  {https://arxiv.org/abs/1701.03168} {arXiv:1701.03168} \BibitemShut {NoStop}%
\bibitem [{\citenamefont {Albert}\ \emph {et~al.}(2018)\citenamefont {Albert}
  \emph {et~al.}}]{HAWC:2018szf}%
  \BibitemOpen
  \bibfield  {author} {\bibinfo {author} {\bibfnamefont {A.}~\bibnamefont
  {Albert}} \emph {et~al.} (\bibinfo {collaboration} {HAWC}),\ }\bibfield
  {title} {\bibinfo {title} {{Constraints on Spin-Dependent Dark Matter
  Scattering with Long-Lived Mediators from TeV Observations of the Sun with
  HAWC}},\ }\href {https://doi.org/10.1103/PhysRevD.98.123012} {\bibfield
  {journal} {\bibinfo  {journal} {Phys. Rev. D}\ }\textbf {\bibinfo {volume}
  {98}},\ \bibinfo {pages} {123012} (\bibinfo {year} {2018})},\ \Eprint
  {https://arxiv.org/abs/1808.05624} {arXiv:1808.05624 [hep-ph]} \BibitemShut
  {NoStop}%
\bibitem [{\citenamefont {Niblaeus}\ \emph {et~al.}(2019)\citenamefont
  {Niblaeus}, \citenamefont {Beniwal},\ and\ \citenamefont
  {Edsjo}}]{Niblaeus:2019gjk}%
  \BibitemOpen
  \bibfield  {author} {\bibinfo {author} {\bibfnamefont {C.}~\bibnamefont
  {Niblaeus}}, \bibinfo {author} {\bibfnamefont {A.}~\bibnamefont {Beniwal}},\
  and\ \bibinfo {author} {\bibfnamefont {J.}~\bibnamefont {Edsjo}},\ }\bibfield
   {title} {\bibinfo {title} {{Neutrinos and gamma rays from long-lived
  mediator decays in the Sun}},\ }\href
  {https://doi.org/10.1088/1475-7516/2019/11/011} {\bibfield  {journal}
  {\bibinfo  {journal} {JCAP}\ }\textbf {\bibinfo {volume} {11}},\ \bibinfo
  {pages} {011}},\ \Eprint {https://arxiv.org/abs/1903.11363} {arXiv:1903.11363
  [astro-ph.HE]} \BibitemShut {NoStop}%
\bibitem [{\citenamefont {Mazziotta}\ \emph {et~al.}(2020)\citenamefont
  {Mazziotta}, \citenamefont {Loparco}, \citenamefont {Serini}, \citenamefont
  {Cuoco}, \citenamefont {De~La Torre~Luque}, \citenamefont {Gargano},\ and\
  \citenamefont {Gustafsson}}]{Mazziotta:2020foa}%
  \BibitemOpen
  \bibfield  {author} {\bibinfo {author} {\bibfnamefont {M.~N.}\ \bibnamefont
  {Mazziotta}}, \bibinfo {author} {\bibfnamefont {F.}~\bibnamefont {Loparco}},
  \bibinfo {author} {\bibfnamefont {D.}~\bibnamefont {Serini}}, \bibinfo
  {author} {\bibfnamefont {A.}~\bibnamefont {Cuoco}}, \bibinfo {author}
  {\bibfnamefont {P.}~\bibnamefont {De~La Torre~Luque}}, \bibinfo {author}
  {\bibfnamefont {F.}~\bibnamefont {Gargano}},\ and\ \bibinfo {author}
  {\bibfnamefont {M.}~\bibnamefont {Gustafsson}},\ }\bibfield  {title}
  {\bibinfo {title} {{Search for dark matter signatures in the gamma-ray
  emission towards the Sun with the Fermi Large Area Telescope}},\ }\href
  {https://doi.org/10.1103/PhysRevD.102.022003} {\bibfield  {journal} {\bibinfo
   {journal} {Phys. Rev. D}\ }\textbf {\bibinfo {volume} {102}},\ \bibinfo
  {pages} {022003} (\bibinfo {year} {2020})},\ \Eprint
  {https://arxiv.org/abs/2006.04114} {arXiv:2006.04114 [astro-ph.HE]}
  \BibitemShut {NoStop}%
\bibitem [{\citenamefont {Choi}\ \emph {et~al.}(2015)\citenamefont {Choi} \emph
  {et~al.}}]{Choi:2015ara}%
  \BibitemOpen
  \bibfield  {author} {\bibinfo {author} {\bibfnamefont {K.}~\bibnamefont
  {Choi}} \emph {et~al.} (\bibinfo {collaboration} {Super-Kamiokande}),\
  }\bibfield  {title} {\bibinfo {title} {Search for neutrinos from annihilation
  of captured low-mass dark matter particles in the {{Sun}} by
  {{Super}}-{{Kamiokande}}},\ }\href
  {https://doi.org/10.1103/PhysRevLett.114.141301} {\bibfield  {journal}
  {\bibinfo  {journal} {Phys. Rev. Lett.}\ }\textbf {\bibinfo {volume} {114}},\
  \bibinfo {pages} {141301} (\bibinfo {year} {2015})},\ \Eprint
  {https://arxiv.org/abs/1503.04858} {arXiv:1503.04858} \BibitemShut {NoStop}%
%%CITATION = ARXIV:1503.04858;%%
\bibitem [{\citenamefont {Aartsen}\ \emph
  {et~al.}(2017{\natexlab{a}})\citenamefont {Aartsen} \emph
  {et~al.}}]{2017EPJC...77..146A}%
  \BibitemOpen
  \bibfield  {author} {\bibinfo {author} {\bibfnamefont {M.~G.}\ \bibnamefont
  {Aartsen}} \emph {et~al.},\ }\bibfield  {title} {\bibinfo {title} {Search for
  annihilating dark matter in the {{Sun}} with 3 years of {{IceCube}} data},\
  }\href {https://doi.org/10.1140/epjc/s10052-017-4689-9} {\bibfield  {journal}
  {\bibinfo  {journal} {Eur. Phys. J. C}\ }\textbf {\bibinfo {volume} {77}},\
  \bibinfo {eid} {146} (\bibinfo {year} {2017}{\natexlab{a}})},\ \Eprint
  {https://arxiv.org/abs/1612.05949} {arXiv:1612.05949} \BibitemShut {NoStop}%
\bibitem [{\citenamefont {Colom~i Bernadich}\ and\ \citenamefont {P\'erez
  de~los Heros}(2020)}]{ColomBernadich:2019sqc}%
  \BibitemOpen
  \bibfield  {author} {\bibinfo {author} {\bibfnamefont {M.}~\bibnamefont
  {Colom~i Bernadich}}\ and\ \bibinfo {author} {\bibfnamefont {C.}~\bibnamefont
  {P\'erez de~los Heros}},\ }\bibfield  {title} {\bibinfo {title} {{Limits on
  Kaluza\textendash{}Klein dark matter annihilation in the Sun from recent
  IceCube results}},\ }\href {https://doi.org/10.1140/epjc/s10052-020-7708-1}
  {\bibfield  {journal} {\bibinfo  {journal} {Eur. Phys. J. C}\ }\textbf
  {\bibinfo {volume} {80}},\ \bibinfo {pages} {129} (\bibinfo {year} {2020})},\
  \Eprint {https://arxiv.org/abs/1912.04585} {arXiv:1912.04585 [astro-ph.HE]}
  \BibitemShut {NoStop}%
\bibitem [{\citenamefont {{Adrian-Martinez}}\ \emph {et~al.}(2016)\citenamefont
  {{Adrian-Martinez}} \emph {et~al.}}]{2016PhLB..759...69A}%
  \BibitemOpen
  \bibfield  {author} {\bibinfo {author} {\bibfnamefont {S.}~\bibnamefont
  {{Adrian-Martinez}}} \emph {et~al.},\ }\bibfield  {title} {\bibinfo {title}
  {Limits on dark matter annihilation in the sun using the {{ANTARES}} neutrino
  telescope},\ }\href {https://doi.org/10.1016/j.physletb.2016.05.019}
  {\bibfield  {journal} {\bibinfo  {journal} {Physics Letters B}\ }\textbf
  {\bibinfo {volume} {759}},\ \bibinfo {pages} {69} (\bibinfo {year} {2016})},\
  \Eprint {https://arxiv.org/abs/1603.02228} {arXiv:1603.02228} \BibitemShut
  {NoStop}%
\bibitem [{\citenamefont {Adri\'an-Mart\'\i{}nez}\ \emph
  {et~al.}(2016)\citenamefont {Adri\'an-Mart\'\i{}nez} \emph
  {et~al.}}]{ANTARES:2016obx}%
  \BibitemOpen
  \bibfield  {author} {\bibinfo {author} {\bibfnamefont {S.}~\bibnamefont
  {Adri\'an-Mart\'\i{}nez}} \emph {et~al.} (\bibinfo {collaboration}
  {ANTARES}),\ }\bibfield  {title} {\bibinfo {title} {{A search for Secluded
  Dark Matter in the Sun with the ANTARES neutrino telescope}},\ }\href
  {https://doi.org/10.1088/1475-7516/2016/05/016} {\bibfield  {journal}
  {\bibinfo  {journal} {JCAP}\ }\textbf {\bibinfo {volume} {05}},\ \bibinfo
  {pages} {016}},\ \Eprint {https://arxiv.org/abs/1602.07000} {arXiv:1602.07000
  [hep-ex]} \BibitemShut {NoStop}%
\bibitem [{\citenamefont {Leane}\ \emph {et~al.}(2018)\citenamefont {Leane},
  \citenamefont {Slatyer}, \citenamefont {Beacom},\ and\ \citenamefont
  {Ng}}]{Leane:2018kjk}%
  \BibitemOpen
  \bibfield  {author} {\bibinfo {author} {\bibfnamefont {R.~K.}\ \bibnamefont
  {Leane}}, \bibinfo {author} {\bibfnamefont {T.~R.}\ \bibnamefont {Slatyer}},
  \bibinfo {author} {\bibfnamefont {J.~F.}\ \bibnamefont {Beacom}},\ and\
  \bibinfo {author} {\bibfnamefont {K.~C.~Y.}\ \bibnamefont {Ng}},\ }\bibfield
  {title} {\bibinfo {title} {Gev-scale thermal wimps: Not even slightly ruled
  out},\ }\href {https://doi.org/10.1103/PhysRevD.98.023016} {\bibfield
  {journal} {\bibinfo  {journal} {Phys. Rev. D}\ }\textbf {\bibinfo {volume}
  {98}},\ \bibinfo {pages} {023016} (\bibinfo {year} {2018})}\BibitemShut
  {NoStop}%
\bibitem [{\citenamefont {Abbasi}\ \emph {et~al.}(2010)\citenamefont {Abbasi}
  \emph {et~al.}}]{IceCube:2010dpc}%
  \BibitemOpen
  \bibfield  {author} {\bibinfo {author} {\bibfnamefont {R.}~\bibnamefont
  {Abbasi}} \emph {et~al.} (\bibinfo {collaboration} {IceCube}),\ }\bibfield
  {title} {\bibinfo {title} {{Calibration and Characterization of the IceCube
  Photomultiplier Tube}},\ }\href {https://doi.org/10.1016/j.nima.2010.03.102}
  {\bibfield  {journal} {\bibinfo  {journal} {Nucl. Instrum. Meth. A}\ }\textbf
  {\bibinfo {volume} {618}},\ \bibinfo {pages} {139} (\bibinfo {year}
  {2010})},\ \Eprint {https://arxiv.org/abs/1002.2442} {arXiv:1002.2442
  [astro-ph.IM]} \BibitemShut {NoStop}%
\bibitem [{\citenamefont {Abbasi}\ \emph {et~al.}(2012)\citenamefont {Abbasi}
  \emph {et~al.}}]{IceCube:2011ucd}%
  \BibitemOpen
  \bibfield  {author} {\bibinfo {author} {\bibfnamefont {R.}~\bibnamefont
  {Abbasi}} \emph {et~al.} (\bibinfo {collaboration} {IceCube}),\ }\bibfield
  {title} {\bibinfo {title} {{The Design and Performance of IceCube
  DeepCore}},\ }\href {https://doi.org/10.1016/j.astropartphys.2012.01.004}
  {\bibfield  {journal} {\bibinfo  {journal} {Astropart. Phys.}\ }\textbf
  {\bibinfo {volume} {35}},\ \bibinfo {pages} {615} (\bibinfo {year} {2012})},\
  \Eprint {https://arxiv.org/abs/1109.6096} {arXiv:1109.6096 [astro-ph.IM]}
  \BibitemShut {NoStop}%
\bibitem [{\citenamefont {Aartsen}\ \emph
  {et~al.}(2017{\natexlab{b}})\citenamefont {Aartsen} \emph
  {et~al.}}]{Aartsen_2017}%
  \BibitemOpen
  \bibfield  {author} {\bibinfo {author} {\bibfnamefont {M.}~\bibnamefont
  {Aartsen}} \emph {et~al.} (\bibinfo {collaboration} {IceCube}),\ }\bibfield
  {title} {\bibinfo {title} {The {IceCube} neutrino observatory:
  instrumentation and online systems},\ }\href
  {https://doi.org/10.1088/1748-0221/12/03/p03012} {\bibfield  {journal}
  {\bibinfo  {journal} {JINST}\ }\textbf {\bibinfo {volume} {12}}\bibinfo
  {number} { (03)},\ \bibinfo {pages} {P03012}}\BibitemShut {NoStop}%
\bibitem [{\citenamefont {Aartsen}\ \emph {et~al.}(2014)\citenamefont {Aartsen}
  \emph {et~al.}}]{Aartsen:2013vja}%
  \BibitemOpen
\bibfield  {number} {  }\bibfield  {author} {\bibinfo {author} {\bibfnamefont
  {M.~G.}\ \bibnamefont {Aartsen}} \emph {et~al.} (\bibinfo {collaboration}
  {IceCube}),\ }\bibfield  {title} {\bibinfo {title} {Energy {{Reconstruction
  Methods}} in the {{IceCube Neutrino Telescope}}},\ }\href
  {https://doi.org/10.1088/1748-0221/9/03/P03009} {\bibfield  {journal}
  {\bibinfo  {journal} {JINST}\ }\textbf {\bibinfo {volume} {9}},\ \bibinfo
  {pages} {P03009}},\ \Eprint {https://arxiv.org/abs/1311.4767}
  {arXiv:1311.4767} \BibitemShut {NoStop}%
%%CITATION = ARXIV:1311.4767;%%
\bibitem [{\citenamefont {Aartsen}\ \emph {et~al.}(2018)\citenamefont {Aartsen}
  \emph {et~al.}}]{PhysRevLett.120.071801}%
  \BibitemOpen
  \bibfield  {author} {\bibinfo {author} {\bibfnamefont {M.~G.}\ \bibnamefont
  {Aartsen}} \emph {et~al.} (\bibinfo {collaboration} {IceCube}),\ }\bibfield
  {title} {\bibinfo {title} {Measurement of atmospheric neutrino oscillations
  at 6--56 gev with icecube deepcore},\ }\href
  {https://doi.org/10.1103/PhysRevLett.120.071801} {\bibfield  {journal}
  {\bibinfo  {journal} {Phys. Rev. Lett.}\ }\textbf {\bibinfo {volume} {120}},\
  \bibinfo {pages} {071801} (\bibinfo {year} {2018})}\BibitemShut {NoStop}%
\bibitem [{\citenamefont {{Gould}}(1987)}]{1987ApJ...321..560G}%
  \BibitemOpen
  \bibfield  {author} {\bibinfo {author} {\bibfnamefont {A.}~\bibnamefont
  {{Gould}}},\ }\bibfield  {title} {\bibinfo {title} {{Weakly Interacting
  Massive Particle Distribution in and Evaporation from the Sun}},\ }\href
  {https://doi.org/10.1086/165652} {\bibfield  {journal} {\bibinfo  {journal}
  {\apj}\ }\textbf {\bibinfo {volume} {321}},\ \bibinfo {pages} {560} (\bibinfo
  {year} {1987})}\BibitemShut {NoStop}%
\bibitem [{\citenamefont {Garani}\ and\ \citenamefont
  {Palomares-Ruiz}(2021)}]{Garani:2021feo}%
  \BibitemOpen
  \bibfield  {author} {\bibinfo {author} {\bibfnamefont {R.}~\bibnamefont
  {Garani}}\ and\ \bibinfo {author} {\bibfnamefont {S.}~\bibnamefont
  {Palomares-Ruiz}},\ }\bibfield  {title} {\bibinfo {title} {{Evaporation of
  dark matter from celestial bodies}}\ }(\bibinfo {year} {2021})\ \Eprint
  {https://arxiv.org/abs/2104.12757} {arXiv:2104.12757 [hep-ph]} \BibitemShut
  {NoStop}%
\bibitem [{\citenamefont {Widmark}(2017)}]{2017JCAP...05..046W}%
  \BibitemOpen
  \bibfield  {author} {\bibinfo {author} {\bibfnamefont {A.}~\bibnamefont
  {Widmark}},\ }\bibfield  {title} {\bibinfo {title} {Thermalization time
  scales for {{WIMP}} capture by the {{Sun}} in effective theories},\ }\href
  {https://doi.org/10.1088/1475-7516/2017/05/046} {\bibfield  {journal}
  {\bibinfo  {journal} {JCAP}\ }\textbf {\bibinfo {volume} {5}},\ \bibinfo
  {eid} {046}},\ \Eprint {https://arxiv.org/abs/1703.06878} {arXiv:1703.06878}
  \BibitemShut {NoStop}%
\bibitem [{\citenamefont {Blennow}\ \emph {et~al.}(2008)\citenamefont
  {Blennow}, \citenamefont {Edsjo},\ and\ \citenamefont
  {Ohlsson}}]{Blennow:2007tw}%
  \BibitemOpen
  \bibfield  {author} {\bibinfo {author} {\bibfnamefont {M.}~\bibnamefont
  {Blennow}}, \bibinfo {author} {\bibfnamefont {J.}~\bibnamefont {Edsjo}},\
  and\ \bibinfo {author} {\bibfnamefont {T.}~\bibnamefont {Ohlsson}},\
  }\bibfield  {title} {\bibinfo {title} {{Neutrinos from WIMP annihilations
  using a full three-flavor Monte Carlo}},\ }\href
  {https://doi.org/10.1088/1475-7516/2008/01/021} {\bibfield  {journal}
  {\bibinfo  {journal} {JCAP}\ }\textbf {\bibinfo {volume} {01}},\ \bibinfo
  {pages} {021}},\ \Eprint {https://arxiv.org/abs/0709.3898} {arXiv:0709.3898
  [hep-ph]} \BibitemShut {NoStop}%
\bibitem [{\citenamefont {Andreopoulos}\ \emph {et~al.}(2010)\citenamefont
  {Andreopoulos} \emph {et~al.}}]{Andreopoulos:2009rq}%
  \BibitemOpen
  \bibfield  {author} {\bibinfo {author} {\bibfnamefont {C.}~\bibnamefont
  {Andreopoulos}} \emph {et~al.},\ }\bibfield  {title} {\bibinfo {title} {{The
  GENIE Neutrino Monte Carlo Generator}},\ }\href
  {https://doi.org/10.1016/j.nima.2009.12.009} {\bibfield  {journal} {\bibinfo
  {journal} {Nucl. Instrum. Meth. A}\ }\textbf {\bibinfo {volume} {614}},\
  \bibinfo {pages} {87} (\bibinfo {year} {2010})},\ \Eprint
  {https://arxiv.org/abs/0905.2517} {arXiv:0905.2517 [hep-ph]} \BibitemShut
  {NoStop}%
\bibitem [{\citenamefont {Honda}\ \emph {et~al.}(2007)\citenamefont {Honda},
  \citenamefont {Kajita}, \citenamefont {Kasahara}, \citenamefont
  {Midorikawa},\ and\ \citenamefont {Sanuki}}]{Honda:2006qj}%
  \BibitemOpen
  \bibfield  {author} {\bibinfo {author} {\bibfnamefont {M.}~\bibnamefont
  {Honda}}, \bibinfo {author} {\bibfnamefont {T.}~\bibnamefont {Kajita}},
  \bibinfo {author} {\bibfnamefont {K.}~\bibnamefont {Kasahara}}, \bibinfo
  {author} {\bibfnamefont {S.}~\bibnamefont {Midorikawa}},\ and\ \bibinfo
  {author} {\bibfnamefont {T.}~\bibnamefont {Sanuki}},\ }\bibfield  {title}
  {\bibinfo {title} {{Calculation of atmospheric neutrino flux using the
  interaction model calibrated with atmospheric muon data}},\ }\href
  {https://doi.org/10.1103/PhysRevD.75.043006} {\bibfield  {journal} {\bibinfo
  {journal} {Phys. Rev. D}\ }\textbf {\bibinfo {volume} {75}},\ \bibinfo
  {pages} {043006} (\bibinfo {year} {2007})},\ \Eprint
  {https://arxiv.org/abs/astro-ph/0611418} {arXiv:astro-ph/0611418}
  \BibitemShut {NoStop}%
\bibitem [{\citenamefont {Amole}\ \emph {et~al.}(2019)\citenamefont {Amole}
  \emph {et~al.}}]{PhysRevD.100.022001}%
  \BibitemOpen
  \bibfield  {author} {\bibinfo {author} {\bibfnamefont {C.}~\bibnamefont
  {Amole}} \emph {et~al.} (\bibinfo {collaboration} {PICO}),\ }\bibfield
  {title} {\bibinfo {title} {Dark matter search results from the complete
  exposure of the pico-60 ${\mathrm{c}}_{3}{\mathrm{f}}_{8}$ bubble chamber},\
  }\href {https://doi.org/10.1103/PhysRevD.100.022001} {\bibfield  {journal}
  {\bibinfo  {journal} {Phys. Rev. D}\ }\textbf {\bibinfo {volume} {100}},\
  \bibinfo {pages} {022001} (\bibinfo {year} {2019})}\BibitemShut {NoStop}%
\bibitem [{\citenamefont {{Adrian-Martinez}}\ \emph {et~al.}(2016)\citenamefont
  {{Adrian-Martinez}} \emph {et~al.}}]{Adrian-Martinez:2016gti}%
  \BibitemOpen
  \bibfield  {author} {\bibinfo {author} {\bibfnamefont {S.}~\bibnamefont
  {{Adrian-Martinez}}} \emph {et~al.} (\bibinfo {collaboration} {ANTARES}),\
  }\bibfield  {title} {\bibinfo {title} {Limits on {{Dark Matter Annihilation}}
  in the {{Sun}} using the {{ANTARES Neutrino Telescope}}},\ }\href
  {https://doi.org/10.1016/j.physletb.2016.05.019} {\bibfield  {journal}
  {\bibinfo  {journal} {Phys. Lett.}\ }\textbf {\bibinfo {volume} {B759}},\
  \bibinfo {pages} {69} (\bibinfo {year} {2016})},\ \Eprint
  {https://arxiv.org/abs/1603.02228} {arXiv:1603.02228} \BibitemShut {NoStop}%
%%CITATION = ARXIV:1603.02228;%%
\bibitem [{\citenamefont {Ackermann}\ \emph {et~al.}(2006)\citenamefont
  {Ackermann} \emph {et~al.}}]{Ackermann:2006pva}%
  \BibitemOpen
  \bibfield  {author} {\bibinfo {author} {\bibfnamefont {M.}~\bibnamefont
  {Ackermann}} \emph {et~al.},\ }\bibfield  {title} {\bibinfo {title} {{Optical
  properties of deep glacial ice at the South Pole}},\ }\href
  {https://doi.org/10.1029/2005JD006687} {\bibfield  {journal} {\bibinfo
  {journal} {J. Geophys. Res.}\ }\textbf {\bibinfo {volume} {111}},\ \bibinfo
  {pages} {D13203} (\bibinfo {year} {2006})}\BibitemShut {NoStop}%
\bibitem [{\citenamefont {Fiedlschuster}(2019)}]{Fiedlschuster:2019unl}%
  \BibitemOpen
  \bibfield  {author} {\bibinfo {author} {\bibfnamefont {S.}~\bibnamefont
  {Fiedlschuster}},\ }\bibfield  {title} {\bibinfo {title} {{The Effect of Hole
  Ice on the Propagation and Detection of Light in IceCube}},\ }\href@noop {}
  {\bibfield  {journal} {\bibinfo  {journal} {ArXiv e-prints}\ } (\bibinfo
  {year} {2019})},\ \Eprint {https://arxiv.org/abs/1904.08422}
  {arXiv:1904.08422 [physics.ins-det]} \BibitemShut {NoStop}%
\bibitem [{\citenamefont {Jungman}\ \emph {et~al.}(1996)\citenamefont
  {Jungman}, \citenamefont {Kamionkowski},\ and\ \citenamefont
  {Griest}}]{Jungman:1995df}%
  \BibitemOpen
  \bibfield  {author} {\bibinfo {author} {\bibfnamefont {G.}~\bibnamefont
  {Jungman}}, \bibinfo {author} {\bibfnamefont {M.}~\bibnamefont
  {Kamionkowski}},\ and\ \bibinfo {author} {\bibfnamefont {K.}~\bibnamefont
  {Griest}},\ }\bibfield  {title} {\bibinfo {title} {Supersymmetric dark
  matter},\ }\href {https://doi.org/10.1016/0370-1573(95)00058-5} {\bibfield
  {journal} {\bibinfo  {journal} {Phys. Rept.}\ }\textbf {\bibinfo {volume}
  {267}},\ \bibinfo {pages} {195} (\bibinfo {year} {1996})},\ \Eprint
  {https://arxiv.org/abs/hep-ph/9506380} {arXiv:hep-ph/9506380} \BibitemShut
  {NoStop}%
%%CITATION = HEP-PH/9506380;%%
\bibitem [{\citenamefont {Edsjo}\ \emph {et~al.}(2017)\citenamefont {Edsjo},
  \citenamefont {Elevant}, \citenamefont {Enberg},\ and\ \citenamefont
  {Niblaeus}}]{Edsjo:2017kjk}%
  \BibitemOpen
  \bibfield  {author} {\bibinfo {author} {\bibfnamefont {J.}~\bibnamefont
  {Edsjo}}, \bibinfo {author} {\bibfnamefont {J.}~\bibnamefont {Elevant}},
  \bibinfo {author} {\bibfnamefont {R.}~\bibnamefont {Enberg}},\ and\ \bibinfo
  {author} {\bibfnamefont {C.}~\bibnamefont {Niblaeus}},\ }\bibfield  {title}
  {\bibinfo {title} {Neutrinos from cosmic ray interactions in the {{Sun}}},\
  }\href {https://doi.org/10.1088/1475-7516/2017/06/033} {\bibfield  {journal}
  {\bibinfo  {journal} {JCAP}\ }\textbf {\bibinfo {volume} {1706}}\bibfield
  {number} {\bibinfo  {number} { (06)},\ \bibinfo {pages} {033}},\ }\Eprint
  {https://arxiv.org/abs/1704.02892} {arXiv:1704.02892} \BibitemShut {NoStop}%
%%CITATION = ARXIV:1704.02892;%%
\bibitem [{\citenamefont {Arg{\"u}elles}\ \emph {et~al.}(2017)\citenamefont
  {Arg{\"u}elles}, \citenamefont {{de Wasseige}}, \citenamefont {Fedynitch},\
  and\ \citenamefont {Jones}}]{2017JCAP...07..024A}%
  \BibitemOpen
  \bibfield  {author} {\bibinfo {author} {\bibfnamefont {C.~A.}\ \bibnamefont
  {Arg{\"u}elles}}, \bibinfo {author} {\bibfnamefont {G.}~\bibnamefont {{de
  Wasseige}}}, \bibinfo {author} {\bibfnamefont {A.}~\bibnamefont
  {Fedynitch}},\ and\ \bibinfo {author} {\bibfnamefont {B.~J.~P.}\ \bibnamefont
  {Jones}},\ }\bibfield  {title} {\bibinfo {title} {Solar atmospheric neutrinos
  and the sensitivity floor for solar dark matter annihilation searches},\
  }\href {https://doi.org/10.1088/1475-7516/2017/07/024} {\bibfield  {journal}
  {\bibinfo  {journal} {JCAP}\ }\textbf {\bibinfo {volume} {7}},\ \bibinfo
  {eid} {024}},\ \Eprint {https://arxiv.org/abs/1703.07798} {arXiv:1703.07798}
  \BibitemShut {NoStop}%
\bibitem [{\citenamefont {Ng}\ \emph {et~al.}(2017)\citenamefont {Ng},
  \citenamefont {Beacom}, \citenamefont {Peter},\ and\ \citenamefont
  {Rott}}]{2017PhRvD..96j3006N}%
  \BibitemOpen
  \bibfield  {author} {\bibinfo {author} {\bibfnamefont {K.~C.~Y.}\
  \bibnamefont {Ng}}, \bibinfo {author} {\bibfnamefont {J.~F.}\ \bibnamefont
  {Beacom}}, \bibinfo {author} {\bibfnamefont {A.~H.~G.}\ \bibnamefont
  {Peter}},\ and\ \bibinfo {author} {\bibfnamefont {C.}~\bibnamefont {Rott}},\
  }\bibfield  {title} {\bibinfo {title} {Solar atmospheric neutrinos: {{A}} new
  neutrino floor for dark matter searches},\ }\href
  {https://doi.org/10.1103/PhysRevD.96.103006} {\bibfield  {journal} {\bibinfo
  {journal} {Phys. Rev. D}\ }\textbf {\bibinfo {volume} {96}},\ \bibinfo {eid}
  {103006} (\bibinfo {year} {2017})},\ \Eprint
  {https://arxiv.org/abs/1703.10280} {arXiv:1703.10280} \BibitemShut {NoStop}%
\bibitem [{\citenamefont {Aartsen}\ \emph {et~al.}(2021)\citenamefont {Aartsen}
  \emph {et~al.}}]{Aartsen_2021}%
  \BibitemOpen
  \bibfield  {author} {\bibinfo {author} {\bibfnamefont {M.}~\bibnamefont
  {Aartsen}} \emph {et~al.},\ }\bibfield  {title} {\bibinfo {title} {Searches
  for neutrinos from cosmic-ray interactions in the sun using seven years of
  {IceCube} data},\ }\href {https://doi.org/10.1088/1475-7516/2021/02/025}
  {\bibfield  {journal} {\bibinfo  {journal} {JCAP}\ }\textbf {\bibinfo
  {volume} {2021}}\bibinfo  {number} { (02)},\ \bibinfo {pages}
  {025}}\BibitemShut {NoStop}%
\bibitem [{\citenamefont {Peters}\ \emph {et~al.}(2021)\citenamefont {Peters},
  \citenamefont {Choi},\ and\ \citenamefont {Nisa}}]{Peters:2021afm}%
  \BibitemOpen
\bibfield  {number} {  }\bibfield  {author} {\bibinfo {author} {\bibfnamefont
  {L.}~\bibnamefont {Peters}}, \bibinfo {author} {\bibfnamefont
  {K.}~\bibnamefont {Choi}},\ and\ \bibinfo {author} {\bibfnamefont {M.~U.}\
  \bibnamefont {Nisa}} (\bibinfo {collaboration} {IceCube}),\ }\bibfield
  {title} {\bibinfo {title} {{Constraining Non-Standard Dark Matter-Nucleon
  Interactions with IceCube}},\ }in\ \href@noop {} {\emph {\bibinfo {booktitle}
  {{37th International Cosmic Ray Conference}}}}\ (\bibinfo {year} {2021})\
  \Eprint {https://arxiv.org/abs/2108.05203} {arXiv:2108.05203 [hep-ex]}
  \BibitemShut {NoStop}%
\end{thebibliography}%
\end{document}